\documentclass[10pt,aps,pra,twocolumn,showpacs,superscriptaddress,nobalancelastpage,notitlepage,longbibliography]{revtex4-2}

\usepackage{graphicx,color,mathbbol,amsthm,amsmath,amsfonts,amssymb,bm,braket,dsfont,footmisc,mathrsfs,multirow,latexsym,times}

\usepackage[normalem]{ulem} 
\usepackage{array}
\usepackage{upgreek}
\usepackage{mathrsfs}
\usepackage{braket}
\usepackage{xcolor}
\usepackage{comment}

\usepackage[colorlinks]{hyperref}
\hypersetup{
 colorlinks=true,
 citecolor=magenta,
 linkcolor=blue,
 urlcolor=cyan}

\newcommand{\erf}[1]{Eq. (\ref{#1})}

\newcommand{\LV}[1]{\textcolor[rgb]{.7,0,0}{#1}}

\newcommand{\expect}[1]{\langle #1 \rangle}

\begin{document}

\title{SPAM-Robust Multi-axis Quantum Noise Spectroscopy in Temporally Correlated Environments}

\author{Muhammad Qasim Khan}
\affiliation{Department of Physics and Astronomy, Dartmouth College, Hanover, New Hampshire 03755, USA}

\author{Wenzheng Dong}
\affiliation{Department of Physics and Astronomy, Dartmouth College, Hanover, New Hampshire 03755, USA}

\author{Leigh M. Norris}
\affiliation{\mbox{Johns Hopkins University Applied Physics Laboratory, 11100 Johns Hopkins Road, Laurel, MD, 20723, USA} }

\author{Lorenza Viola}
\affiliation{Department of Physics and Astronomy, Dartmouth College, Hanover, New Hampshire 03755, USA}

\date{\today}

\begin{abstract}
Characterizing temporally correlated (``non-Markovian'') noise is a key prerequisite for achieving noise-tailored error mitigation and optimal device performance. Quantum noise spectroscopy can afford quantitative estimation of the noise spectral features; however, in its current form it is highly vulnerable to implementation non-idealities, notably, state-preparation and measurement (SPAM) errors. Further to that, existing protocols have been mostly developed for dephasing-dominated noise processes, with competing dephasing and relaxation effects being largely unaccounted for. We introduce quantum noise spectroscopy protocols inspired by spin-locking techniques that enable the characterization of arbitrary temporally correlated multi-axis noise on a qubit with fixed energy splitting, while remaining resilient to realistic static SPAM errors. By validating our protocol's performance in both numerical simulation and cloud-based IBM quantum processors, we demonstrate the successful separation and estimation of native noise spectrum components as well as SPAM error rates. We find that SPAM errors can significantly alter or mask important noise features, with spectra overestimated by up to 26.4\% in a classical noise regime. Clear signatures of non-classical noise are manifest in the reconstructed IBM-qubit dephasing spectra, once SPAM-error effects are compensated for. Our work provides a timely tool for benchmarking realistic sources of noise in qubit devices.
\end{abstract}

\maketitle


\section{Introduction}

Reliably characterizing the noise that limits qubit-device performance can provide important physical insight toward improving qubit design and material properties \cite{IrfanReview}, and is a prerequisite for developing efficient noise-adapted and noise-optimized quantum control protocols and error mitigation strategies \cite{Layden,Fadi,Wenzheng,Flammia,Wagner, PettaNonMarkovian,PhysRevLett.131.210802}. Thanks to their low experimental overhead, circuit-level characterization techniques based on randomized benchmarking \cite{Emerson_2005,PhysRevA.77.012307,Wallman_2014} have become a method-of-choice for understanding how noise impacts average gate performance, with more complex gate-set tomography protocols \cite{Nielsen2021gatesettomography,CompressedGST} offering instead a complete tomographic description of the implementation of an entire set of quantum gates. While these methods share the important practical advantage of incorporating robustness against {\em state preparation and measurement (SPAM)} errors, they have been developed and largely employed thus far under the assumption that the underlying noise is Markovian, resulting in error processes that are probabilistic in nature and uncorrelated in time. 

Realistic multi-qubit noise processes generically entail both temporal and spatial correlations, however. In addition, they may originate from genuinely non-classical (non-commuting) noise environments. Due to the variety of microscopic mechanisms that solid-state hosts can engender (e.g., local electric and magnetic field fluctuations, coupling to lattice nuclear spins, two-level defects or phonons), temporally correlated noise is especially prominent in superconducting and spin qubits. The ensuing {\em non-Markovian} dynamics can engender unwanted history dependence, crosstalk and error propagation which, if unaccounted for, can pose a serious challenge to scalability and, ultimately, quantum fault-tolerance \cite{Sarovar2020detectingcrosstalk,NgPreskill,PhysRevA.104.L050404, Michiel}. While progress is being made in accounting for the effects of temporal correlations in the context of both randomized benchmarking protocols \cite{Ball2016,Mike,Romero2021} and quantum process \cite{PettaNonMarkovian} and process tensor tomography approaches \cite{White2020,NonMarkovianGST}, the extra resource requirements are significant \cite{White2022}. As a result, methods for characterizing noise correlations and non-Markovian error behavior presently remain a focus of active investigation.

Hamiltonian-level {\em Quantum Noise Spectroscopy} (QNS) protocols are, in principle, ideally suited for providing the information about the temporal and spatial noise correlations that is needed to achieve noise-optimized prediction and gate design in non-Markovian settings. The core principle of QNS is the fact that the spectral response of a quantum sensor (a single qubit in the simplest case) can be suitably shaped through the application of external control \cite{Schoelkopf2003,Faoro}, in such a way that information about the relevant frequency-domain noise spectra (or their time-domain correlation functions) may be inferred from access to system-only observables. This is achieved through an appropriate ``deconvolution'' procedure, without {\em a priori} assuming known, parametrized functional forms for the target estimates. By now, a variety of ``non-parametric'' QNS approaches have been developed, differing in both the complexity of the underlying estimation problem and the operating control modalities, with successful experimental validations being reported across several experimental platforms. In particular, multi-pulse single-qubit QNS protocols inspired to dynamical decoupling \cite{PhysRevLett.107.230501,yuge2011measurement,Bylander2011Jul, PhysRevA.86.012314,qns,chan2018assessment,williams2023quantifying} have been extended to allow characterization of dephasing noise in a non-Gaussian regime \cite{PhysRevLett.116.150503,Sung2019,Ramon} and in a multi-qubit setting \cite{PhysRevA.95.022121} -- revealing strong temporal and spatial noise cross-correlations in silicon spin-qubit devices \cite{Tarucha1,Tarucha2}. Likewise, QNS protocols inspired by spin-locking (SL) ``$T_{1\rho}$ relaxometry'' \cite{Yan2013} have been used to sample general non-classical noise spectra \cite{Yan2018} and to obtain a complete characterization of spatial noise correlations in a superconducting two-qubit device, including non-classical cross-correlation spectra \cite{Uwe2020}. The use of optimally band-limited control modulation based on ``Slepian functions''  has enabled the design of QNS schemes for characterizing time-correlated control noise \cite{Frey2017,Frey2020,Maloney}, while a ``control-free'' Fourier-transform QNS approach \cite{Castillo} and a resource-efficient characterization based on ``frames'' \cite{PazSilvaFrames,Wenzheng,wang2022digital} have also been recently proposed.
  
Notwithstanding the above advances, even at the single-qubit level existing QNS protocols still suffer from two important limitations. First, they have been overwhelmingly developed and experimentally validated in regimes where ambient noise is assumed to be dominated by dephasing processes (``single-axis noise''), with the characterization of competing dephasing and relaxation effects (``multi-axis noise'') being largely unaccounted for. More specifically, single-qubit multi-axis QNS protocols based on pulsed control have been put forward in \cite{CombMultiaxis}. However, due to limitations stemming from the  ``frequency-comb'' approach they rely upon to sample the target spectra, these protocols do not lend themselves to characterizing spectra with wide support in frequency space -- as typically arising for relaxation phenomena. For superconducting qubits, SL QNS was originally employed to reconstruct the spectra of both the tunnel-coupling and flux noise responsible for dephasing ($T_2$) and relaxation ($T_1$) processes \cite{Yan2013}, and a characterization of $T_1$ noise was also later reported in \cite{FluxQRevisited}. However, the protocols were either leveraging the ability to effectively turn-off a source of noise and independently characterize the other, or to tune the energy splitting of the probing qubit -- therefore limiting portability to different qubit architectures. Second, QNS protocols are highly vulnerable to SPAM errors in their current form. While SPAM errors do not accumulate with circuit depth, they are a significant (often dominant) source of error in NISQ processors, with SPAM error rates between 3\%-10\% being reported, for instance, for both Google and IBM processors \cite{Arute2019,Barber2023postselectionfree,PhysRevA.103.042605,PRXQuantum.2.040326}. This can overwhelm the target spectral signal and significantly bias spectral reconstructions, particularly in the relevant scenario where weak native noise sources are targeted for estimation.
 
In this paper, we theoretically develop and validate, through both numerical and cloud-based IBM Quantum Platform (IBM Q) implementations, a single-qubit generalized SL QNS protocol that can characterize arbitrary multi-axis weak noise, and is insensitive to realistic (static) SPAM errors. More precisely, without any prior knowledge about SPAM errors, our protocol is fully insensitive to SPAM in a regime where noise may be assumed to be classical; for general non-classical noise, additional knowledge about either the structure of the SPAM errors or prior characterization of individual state preparation and measurement error rates is needed. While non-tomographic protocols able to separately quantify SPAM contributions have been recently proposed \cite{Laflamme2021,Wei2023}, they rely upon data-intensive Bayesian techniques \cite{Mattias} or access to an ancillary qubit; in line with the fact that measurement errors play a dominant role in current NISQ devices \cite{PhysRevA.100.052315,PhysRevA.103.042605,Geller_2020}, our complete single-qubit protocol will be developed here under the assumption of perfect state preparation. Importantly, our protocols are {\em platform-independent} and do not require, in particular, frequency-tunable qubits. Central to the design of our multi-axis QNS protocol is the representation of the noise spectra in terms of their {\em spherical} components \cite{CombMultiaxis}, as opposed to standard Cartesian components. While requiring the use of both transverse and longitudinal drives relative to the qubit quantization axis, the protocol we present provides a complete SPAM-robust spectral characterization of the noise entering the qubit dynamics.

The content is organized as follows. In Sec.\,\ref{sec::background}, we introduce the relevant open-system model of a driven single qubit sensor in the presence of arbitrary, additive multi-axis noise, which we represent in the spherical basis. Within a weak-coupling approximation, we show how the relevant qubit dynamics may be described in terms of a non-Markovian, time-convolutionless (TCL) master equation (ME) where noise effects are captured at the Gaussian level, up to the second order. By way of grounding our subsequent discussion, we then review the standard SL QNS approach in the presence of a purely dephasing noise environment, in a form suitable for further generalization, and introduce the SPAM error model relevant to the discussion. In Sec.\,\ref{sec:srqnsdephasing}, we quantify the impact of SPAM errors in this standard SL setting, and propose a modified protocol capable to deliver SPAM-free estimates of the target dephasing spectra. The benefits of the SPAM-robust single-axis protocol are demonstrated numerically, by implementing a simulated non-commuting dephasing bath. Sections\,\ref{sec::slqnsmultiaxis} and \ref{sec::srqnsmultiaxis} tackle the general multi-axis noise setting and contain our main results. Specifically, in Sec.\,\ref{sec::slqnsmultiaxis} we leverage a suitable secular approximation to obtain a simplified TCL ME applicable under arbitrary control over time scales long with respect to the inverse qubit energy splitting. By specializing the control to continuous-drives along fixed axes, we construct a generalized SL protocol that can simultaneously estimate all the relevant dephasing and qubit excitation/de-excitation spherical spectra. A SPAM-robust modification of this multi-axis protocol is developed in Sec.\,\ref{sec::srqnsmultiaxis}, and validated on an IBM Q single-qubit device. After discussing the main limitations faced in the implementation and data-analysis, the reconstructed spectra and inferred SPAM parameters are presented. SPAM errors are found to have a significant effect, causing spectra to be generally over-estimated and  introducing unwanted artifacts in the corresponding reconstructions. Section\,\ref{sec::discussion} concludes with an overall assessment and an outlook to future directions. Full detail about the derivation of the relevant TCL ME is given in Appendix\,\ref{appendix::tclmema}, including an independent derivation through multi-scale perturbation theory. The remaining appendices discuss additional technical aspects related, in particular, to the numerical simulation of a non-classical dephasing model (Appendix\,\ref{quantnoisesim}) and the reported IBM Q implementation (Appendix\,\ref{app:expt}).


\section{Background}
\label{sec::background}

\subsection{Open quantum system and control setting} 

\subsubsection{Driven qubit dynamics under multi-axis noise}

A qubit coupled to an arbitrary noisy environment and driven by arbitrary time-dependent open-loop control may be described in the lab frame by a Hamiltonian of the form
\begin{align}
    H_{\text{lab}}(t) &= H_S + H_{\text{ctrl, lab}}(t) + H_{SB}(t) + H_B,
    \label{eq:totHam}
\end{align}
where $H_S \equiv \tfrac{1}{2} \omega_q \sigma_z$ is the qubit Hamiltonian, and $H_{B}$ and $H_{SB}$ are the bath and the system-bath coupling Hamiltonians, respectively. Importantly, the qubit energy splitting $\omega_q $ is assumed to be known but {\em fixed} (non-tunable), and the control Hamiltonian $H_{\text{ctrl, lab}}(t)$ acts non-trivially only on the system. Let $\sigma_u$, $u\in\{x,y,z\}$ denote, as usual, Pauli matrices and, for simplicity, let us for now choose units where $\hbar=1$. By moving to the interaction picture with respect to the free bath Hamiltonian, the coupling Hamiltonian becomes 
\begin{align}
    H_{SB}(t) = \!\!\!\sum_{u=x,y,z}\!\!\!\sigma_u \otimes B_u(t) , \;\; B_u(t) \equiv 
     \widetilde{B}_u(t) +\beta_u(t) \mathbb{1}_B. 
     \label{eq:SystemBath}
\end{align}
Here, $B_u(t)$ are time-dependent (Hermitian) bath operators, which may generally include a quantum (non-commuting) component $\widetilde{B}_u(t)$ and a classical ($c$-number) stochastic process $\beta_u(t) \mathbb{1}_B$, with $\mathbb{1}_B$ being the identity on the bath. Physically, \emph{relative to the qubit energy eigenbasis}, the ``on-axis'' coupling operator along $z$ is responsible for energy-conserving processes contributing only to the (lab-frame) transverse relaxation time, $T_2$, whereas the ``off-axis'' coupling operators along $x, y$ are responsible for both an energy-non-conserving contribution to $T_2$ and to the (lab-frame) longitudinal relaxation time $T_1$. 

Since the presence of off-axis noise induces transitions between the qubit eigenstates, it is convenient to re-express the interaction Hamiltonian in terms of \emph{spherical} as opposed to Cartesian components \cite{CombMultiaxis}. By letting 
\begin{align}
v_{\pm 1}(t)\equiv \frac{v_{x}(t)\pm iv_{y}(t)}{\sqrt{2}}, \quad v_{0}(t)\equiv v_{z}(t), 
\label{eq::spher}
\end{align}
and moving to a frame co-rotating with the qubit frequency, we can rewrite the total  Hamiltonian as
\begin{align}
\label{eq::HRotatingMulti}
    H_{\text{rot}}(t)=\!\!\!\sum_{\alpha=-1,0,1}e^{i\alpha\omega_q t}\sigma_\alpha\otimes B_{-\alpha}(t)+H_\text{ctrl}(t), 
\end{align}
where now $B^\dagger_\alpha(t) = B_{-\alpha}(t)$ and $H_\text{ctrl}(t)$ is 
an appropriately transformed control Hamiltonian. Finally, the action of the applied control is most naturally analyzed by effecting a frame transformation to the interaction picture associated to $H_\text{ctrl}(t)$ (also often referred to as the ``toggling frame''). The relevant open-system Hamiltonian may thus be expressed in the form  
\begin{align}
\label{eq::HTogglingMulti}
\widetilde{H}(t)=\!\!\!\sum_{\alpha,\beta=-1, 0, 1} \!\! y_{\alpha\beta}(t) e^{i\alpha \omega_qt} \sigma_\beta \otimes B_{-\alpha}(t), 
\end{align}
with the {\em control switching functions} $y_{\alpha\beta}(t)$ being, in general, complex scalars defined component-wise via 
\begin{align*}
y_{\alpha, \beta}(t)& \equiv \tfrac{1}{2}\text{Tr}[\sigma_{-\beta}U_\text{ctrl}(t)^\dag\sigma_{\alpha}\,U_\text{ctrl}(t)], 
\end{align*}
in terms of the time-ordered ($\mathcal{T}_+$) control propagator 
$U_\text{ctrl}(t)\equiv \mathcal{T}_+ e^{-i\int_0^tds\,H_\text{ctrl}(s)}.$ 

In order to describe the reduced qubit dynamics, it is necessary in general to introduce suitable approximations that may enable a perturbative treatment. Throughout this work, we will assume that (i) the initial ($t= t_0\equiv0$) system-bath state is factorized, $\rho_{SB}(0) \equiv \rho (0) \otimes \rho_B$, (ii) the initial bath state is stationary, $[\rho_B,H_B]=0$, (iii) the noise process is zero-mean, $\langle B_\alpha(t)\rangle =0, \forall t$, and (iv) the system-bath coupling is sufficiently weak for the effects of $H_{SB}(t)$ (hence $\widetilde{H}(t)$ in Eq.\,\eqref{eq::HTogglingMulti}) to be included only up to the second order (Born approximation). It is important to note that \emph{no} Markovian approximation is made, however. Explicitly, expectation values with respect to the quantum and classical noise components are defined by $\braket{\cdot}\equiv {\mathbb E}\{ {\mathrm{Tr}}_B [\cdot \rho_B] \}$, where ${\mathbb E}\{ \cdot\}$ denotes a classical ensemble average over noise realizations. Under these assumptions, it is possible to employ a standard second-order time-convolutionless (TCL) master equation (ME) \cite{Breuer2007,Uwe2020,PhysRevA.95.022121} which, in the above-defined toggling frame, takes the form
\begin{align}
\label{eq::TCL2ndOrder}
   \dot{\rho}(t)= 
    -\!\int_0^t\!\!ds\,    
    \Big \langle
    \big[\widetilde{H}(t),\big[\widetilde{H}(s),\rho(t)\otimes\rho_B\big]\big]
    \Big \rangle .
\end{align}
This integro-differential equation provides the starting point for the second-order TCL ME we will use to model the SL dynamics of interest. The state operator $\rho(t)$ of the system in Eq.\,(\ref{eq::TCL2ndOrder}) is also defined in the toggling-frame, and will remain so unless otherwise stated.

As a next step, we show how a frequency space representation makes the noise spectral properties explicit.

\subsubsection{Frequency representation: Noise spectra}

In the time domain, the statistical features of the noise are completely encapsulated by the correlation functions of the bath operators. While, in general, the noise process corresponding to $\{B_{\alpha}(t), \rho_B\}$ may be non-Gaussian (hence with non-factorizable higher-order correlation functions), the expansion of the TCL generator to the second order under the weak-coupling approximation in Eq.\,\eqref{eq::TCL2ndOrder} ensures that only the means $\braket{B_\alpha(t)}$ and the two-point correlation functions $\braket{B_{\alpha}(t_1)B_{\beta}(t_2)}$ are relevant in determining the system's dynamics. By assumption (iii), we have $\braket{B_\alpha(t)}=0$. Furthermore, the stationarity assumption (ii) implies time-translational invariance, in the sense that $$\braket{B_{\alpha}(t_1)B_{\beta}(t_2)}=\braket{B_{\alpha}(t_1-t_2)B_{\beta}(0)} \equiv \braket{B_{\alpha}(\tau)B_{\beta}(0)}.$$  
The dependence of the two-point correlation function upon a single time-lag variable, $\tau = t_1-t_2$, makes it then natural (under suitable regularity assumptions) to describe the noise properties in the frequency domain. 

Frequency-domain QNS aims to characterize the spectral properties of noise through orchestrated controlled qubit dynamics. The QNS protocol we introduce here estimates the spherical power spectral densities ({\em spherical spectra} for short), which are Fourier transforms of two-point correlation functions, namely \cite{CombMultiaxis}, 
\begin{align}
\label{eq::spec}
    S_{\alpha,\beta}(\omega)&=\int d\tau \braket{B_\alpha (\tau) B_\beta (0)} e^{-i\omega\tau}.
\end{align}
Since the product of two bath operators can always be decomposed into a sum of a symmetrized and an anti-symmetrized product, in the frequency domain we can similarly distinguish a commuting (``classical'') and a non-commuting (``quantum'') noise contribution to the spectrum by letting \cite{ClerkRMP,PhysRevA.95.022121,CombMultiaxis}
\begin{align*}
    S_{\alpha,\beta}(\omega)\equiv \frac{1}{2}\left[S^+_{\alpha,\beta}(\omega)+S^-_{\alpha,\beta}(\omega)\right].
\end{align*}
That is, if $\{A,B\} \equiv AB +BA$ denotes the anti-commutator between two operators, we have 
\begin{align}
\label{eq::specquant}
    S^+_{\alpha,\beta}(\omega)&=\int d\tau \braket{ \big\{ B_\alpha (\tau), B_\beta (0) \big\}} e^{-i\omega\tau}, \\
\label{eq::specclass}
   S^-_{\alpha,\beta}(\omega)&=\int d\tau \braket{\big[B_\alpha (\tau), B_\beta (0)\big]} e^{-i\omega\tau}.
\end{align}
Equivalently, we may write 
\begin{align}
    S^+_{\alpha,\beta}(\omega)&=S_{\alpha,\beta}(\omega)+S_{\beta,\alpha}(-\omega),
 \label{eq::specquant2}  \\ 
    S^-_{\alpha,\beta}(\omega)&=S_{\alpha,\beta}(\omega)-S_{\beta,\alpha}(-\omega),
 \label{eq::specclass2}   
\end{align}
which also makes the following symmetry properties explicit: 
$[S^{\pm}_{\alpha,\beta}(\omega)]^*=S^{\pm}_{-\alpha,-\beta}(\omega)$. If the noise is purely classical, so that the bath operators commute, $S^-_{\alpha,\beta}(\omega) \equiv 0$ and, in particular, the ``self-spectra'' $S_{\alpha,\alpha}(\omega) = \tfrac{1}{2}S^+_{\alpha,\alpha}(\omega)$ are always even functions of frequency.
Non-zero quantum spectra \mbox{$S^-_{\alpha,\beta}(\omega) \ne 0$} can arise  \textit{only} when the bath operators are non-commuting and, in the setting of linear response, can be directly related to non-vanishing {\em bath susceptibilities} \cite{Gavish,ClerkRMP}. In this case, the quantum self-spectra are always anti-symmetric in the frequency variable, $S^-_{\alpha,\alpha}(\omega)= -S^-_{\alpha,\alpha}(-\omega)$, and ``cross-spectra'' $S^-_{\alpha,\beta \ne\alpha}(\omega)$ control the energy exchange between the system and the bath, ultimately determining frequency-dependent absorption and emission rates \cite{CombMultiaxis}. 

The set of spherical spectra $\{S_{\alpha,\beta} (\omega)\},$ $\alpha, \beta \in\{-1,0,1\}$, completely characterize the statistical properties of the noise that enter the qubit dynamics in the weak-coupling regime. Using the defining relations in Eq.\,\eqref{eq::spher} and the symmetry properties of the spectra, the spherical spectra are found to be related to the {\em Cartesian spectra} \cite{CombMultiaxis} associated with the bath operators in \erf{eq:SystemBath} as follows:
\begin{align*}
    S_{0,0}(\omega)&=S_{zz}(\omega),\\
    S_{\mp1,\pm1}(\omega)&=S_{xx}(\omega)+S_{yy}(\omega)\pm i[S_{xy}(\omega)-S_{yx}(\omega)].
\end{align*}
The goal of QNS is then to obtain accurate estimates for all the dynamically relevant (spherical) spectra.  We will use the qubit dynamics described in Eq.\,\eqref{eq::TCL2ndOrder} under an appropriate control Hamiltonian to do so.

\subsection{Spin-locking QNS under dephasing noise}
\label{SLQNSdephasing}

In preparation for extending SL QNS to multi-axis qubit noise, it is useful to review the standard setting where the noise couples only along the quantization axis. In this case, $B_{+1}(t)=B_{-1}(t)\equiv 0$ and, in the absence of control, $H_{SB}(t) =\sigma_z \otimes B_z(t)$ generates pure dephasing qubit dynamics. The corresponding frequency spectrum can then be simply denoted by $S_{zz}(\omega) \equiv S(\omega)= \tfrac{1}{2}[S^+(\omega) + S^-(\omega)]$. In the important case where the noise environment satisfies a Kubo-Martin-Schwinger condition -- notably, if $\rho_B$ is an  equilibrium Gibbs state with inverse temperature $\beta$, the classical and quantum components of $S(\omega)$ are constrained by the fluctuation-dissipation theorem  \cite{ClerkRMP,Sarracino,PhysRevB.101.155412}, resulting in $S^+(\omega) = \coth(\beta \omega/2) \, S^-(\omega)$.

In a SL protocol, we assume that, in the lab frame, an $x$-polarized coherent drive is applied, of the form $H_{\text{ctrl, lab}}(t) = \Omega (t) \cos(\omega_d t + \phi (t))\sigma_x,$ with the amplitude $\Omega(t)$ and the phase $\phi(t)$ being tunable parameters. If the drive frequency is set to be resonant with the qubit frequency, $\omega_d=\omega_q$, moving to the qubit rotating frame and invoking the rotating-wave approximation to remove terms oscillating at frequency $2\omega_d$ yields the transformed control Hamiltonian
\begin{align*}
    H_{\text{ctrl}}(t) = \tfrac{\Omega (t) }{2}  \big[\cos( \phi (t))\sigma_x + \sin( \phi (t))\sigma_y\big], 
\end{align*}
with $|\Omega | \ll \omega_d$. While modified SL sequences can be considered to improve robustness against control errors \cite{Yan2013}, the basic SL protocol is obtained by specializing the above $  H_{\text{ctrl}}(t)$ to a time-independent $x$-drive. By fixing the phase to $\phi(t)=0$ or $\phi(t)=\pi$, and the amplitude to a constant $\Omega(t)=\Omega$ that can take negative and positive values and the sign of $\Omega$ corresponds to the phase of the drive, we simply have $ H_{\text{ctrl}}(t)= \tfrac{1}{2}\Omega \sigma_x$. The total Hamiltonian in the rotating frame, under this control, takes the form 
\begin{align*}
    H_{\text{rot}}(t) = \tfrac{\Omega}{2}\sigma_x + \sigma_z \otimes B_z(t), 
\end{align*}
which may be interpreted in terms of a new, ``dressed'' qubit whose quantization axis has been rotated by the drive from $z\mapsto x$, with angular-frequency splitting equal to the Rabi frequency, and subject to purely transverse noise. With respect to the effective quantization axis along $\sigma_x$, $B_z(t)$ generates both relaxation and dephasing, often described in terms of characteristic time scales $T_{1\rho}$ and $T_{2\rho}$, respectively \cite{Yan2013}. If the qubit is initially prepared in an eigenstate of $\sigma_x$, it will remain aligned (``locked'') with the constant driving field in the absence of noise -- corresponding to the ``SL condition'' \cite{Yan2013}.
 
As noted, the qubit dynamics may be most easily evaluated by moving to the toggling frame of Eq.\,\eqref{eq::TCL2ndOrder}. For the constant SL drive along $x$, the  toggling frame Hamiltonian is given by
\begin{align*}
\widetilde{H}(t) &= 2[\cos (\Omega t)\sigma_z + \sin (\Omega t)\sigma_y]\otimes B_z(t).
\end{align*}
While it is possible to derive an exact form of the second-order TCL, we use a simplified form of the ME, which holds at sufficiently {\em long evolution times}, such that $|\Omega| t \gg 1$ (see Appendix \ref{mesolve1axis} for additional detail). In this limit, assuming that the noise spectrum $S(\omega)$ varies sufficiently slowly about $\omega =\pm \Omega$, the second-order TCL reduces to
\begin{eqnarray*}
    \dot{\rho}(t)&=&S(-\Omega) \big(\sigma^+ \!\rho(t) \sigma^- \!-\tfrac{1}{2} \{ \sigma^-\sigma^+, \rho(t)\}\big)\\
    &+&S(\Omega) \big(\sigma^- \!\rho(t) \sigma^+ -\!\!\tfrac{1}{2} \{ \sigma^+\sigma^-, \rho(t)\}\big). 
\end{eqnarray*}
Let us introduce the following general notation for eigenstates of Pauli operators in any given frame \footnote{While in previous work \cite{Yan2013, Uwe2020}, different notations (e.g., $\{ \sigma_{X,Y,Z}\}$ or $\{\tau_{x,y,z}\}$) were used to denote Pauli operators in different qubit frames or bases, maintaining a single notation $\{\sigma_{x,y,z}\}$, with the meaning implicit from context, is more natural in the presence of multiaxis noise as we consider.}:
\begin{equation}
\sigma_u \ket{u_\pm}=\pm \ket{u_\pm }, \quad u=x,y,z.
\label{eigs}
\end{equation}
The raising and lowering operators above are defined as $\sigma^\pm\equiv \ket{x_\pm}\bra{x_\mp}$,  where $\ket{x_\pm}$ are eigenstates of $\sigma_x$, often denoted as $\{\ket{+},\ket{-}\}$. In this notation, the computational basis $\{\ket{0},\ket{1}\}$ coincides with $\{\ket{z_{\pm}}\}$. The evolution of the qubit density matrix elements in the toggling frame is governed by the following differential equations (DEs),
\begin{align}
\label{slqnssym1}
    \dot{\rho}_{x_+x_+}(t) &= -S(-\Omega)\rho_{x_+x_+}(t)+S(\Omega)\rho_{x_-x_-}(t),\\
\label{slqnssym2}
    \dot{\rho}_{x_+x_-}(t) & = -\frac{S^+(\Omega)}{2}\rho_{x_+x_-}(t), 
\end{align}
with $\dot{\rho}_{x_-x_-}(t) = - \dot{\rho}_{x_+x_+}(t)$ and $\dot{\rho}_{x_-x_+}(t) = \dot{\rho}_{x_+x_-}(t)^*$ from the requirements of trace preservation and Hermiticity.

While the toggling-frame is a convenient mathematical tool, qubit observables are always measured in the (physically accessible) rotating frame. For purely dephasing noise, by transforming back to the rotating frame and working under the same long-time assumption stated above, we have 
\begin{eqnarray*}
\dot{\rho}_{\text{rot}}(t) =\! &-&i \,[\tfrac{\Omega}{2}\sigma_x, \rho_{\text{rot}}(t)] \\
&+&S(-\Omega) \big(\sigma^+ \!\rho_{\text{rot}}(t) \sigma^- \!-\tfrac{1}{2} \{ \sigma^-\sigma^+, \rho_{\text{rot}}(t)\}\big) \\
&+&S(\Omega) \big(\sigma^- \!\rho_{\text{rot}}(t) \sigma^+ -\!\!\tfrac{1}{2} \{ \sigma^+\sigma^-, \rho_{\text{rot}}(t)\}\big).
\end{eqnarray*} 
It is easy to verify that, since the basis $\{\ket{x_\pm}\}$ diagonalizes $U_\text{ctrl}(t)$, the evolution of the qubit populations given by \erf{slqnssym1} is the same in both the rotating and the toggling frames, that is, $\rho_{x_\pm x_\pm}(t)=\rho_{\text{rot, }x_\pm x_\pm}(t)$. The qubit coherence elements, on the other hand, transform as $$\rho_{x_\pm x_\mp}(t)=\rho_{\text{rot }, x_\pm x_\mp}(t)e^{\pm i \Omega t}, \quad \forall t.$$

For an initial state $\rho(0)$ and drive amplitude $\Omega$, we can determine the expected value of $\sigma_x$ at time $t$ by solving \erf{slqnssym1}, 
\begin{align}
\langle\sigma_x(t)\rangle^\Omega_{\rho(0)} \equiv&\,  \rho_{x_+x_+}(t)-\rho_{x_-x_-}(t)\notag \\ 
=&\frac{e^{-S^+(\Omega)t}}{S^+(\Omega)}\Big[S^+(\Omega) (\rho_{x_+x_+}(0) -\rho_{x_-x_-}(0))\notag\\
\label{eq::ExpectX}
&\hspace*{13mm}+S^-(\Omega) (e^{S^+(\Omega)t}-1)\Big] .
\end{align} 
By evaluating  Eq.\,\eqref{eq::ExpectX} for initial states $\rho(0)=|x_+\rangle\langle x_+|$ and $\rho(0)=|x_-\rangle\langle x_-|$, we obtain expressions for the quantum and classical spectra at drive amplitude $\Omega$,
\begin{eqnarray}
\label{classtheory}
    S^+(\Omega) &\!=\! &\frac{1}{T}\ln\bigg[\frac{2}{\braket{\sigma_x(T)}^\Omega_{x_+}-\braket{\sigma_x(T)}^\Omega_{x_-}}\Bigg],\\
\label{quanttheory}
    S^-(\Omega) &\!=\!& \frac{\braket{\sigma_x(T)}^\Omega_{x_+}+\braket{\sigma_x(T)}^\Omega_{x_-}}{2 \,(1-e^{-S^+(\Omega)T}) } \,S^+(\Omega).\qquad     
\end{eqnarray}
This shows that we can estimate the quantum and classical spectra at frequency $\Omega$ by measuring $\braket{\sigma_x(T)}^\Omega_{x_\pm}$. In practice, the range of frequencies accessible for characterization via SL QNS depends upon various device properties and limitations. An upper limit on $\Omega$ is imposed by considerations related to the need to minimize strong-drive heating. On the other hand, the locking condition is known to break down (e.g., due to system inhomogeneities) when driving is too close to the DC limit ($\Omega = 0$), imposing a lower limit \cite{Yan2013}.

This standard SL QNS protocol is summarized in Table\,\ref{Protocol1}. 

\noindent
\begin{table}
\setlength{\fboxrule}{0.7pt}
\fbox{
\begin{minipage}[t][6.2cm]{0.45\textwidth} 
\raggedright 
        \mbox{\bf{Protocol 1: Single-axis spin-locking QNS} }\\
        \rule{8cm}{0.5pt} 
        \noindent \mbox{\textit{Input.} Driving amplitude $\Omega$. }\\
        \vspace*{1mm}
        \noindent \mbox{\textit{Output.} Dephasing noise spectra $S^\pm(\omega)$ at frequency $\omega=\Omega$.} \\     
        \vspace*{1mm}   
        \noindent \mbox{\textbf{The protocol:} \hspace*{2cm}  }          
\begin{enumerate}
   \item[{\bf (1)}] Prepare the initial state $\ket{x_+}\bra{x_+}$, apply a resonant drive along $x$ with amplitude $\Omega$, let the qubit evolve for time $T$ such that $|\Omega| T \gg 1$, and measure $\sigma_x$.
    \item[{\bf (2)}] Repeat {\bf (1)} sufficiently many times to estimate $\braket{\sigma_x(T)}^\Omega_{x_+}$.
     \item[{\bf (3)}] Repeat steps \textbf{(1)} and \textbf{(2)} with the qubit prepared in $\ket{x_-}\bra{x_-}$ to estimate $\braket{\sigma_x(T)}^\Omega_{x_-}$.
    \item[{\bf (4)}] Estimate $S^\pm(\Omega)$ using Eqs.\,\eqref{classtheory}-\eqref{quanttheory} and the expectation values obtained in steps \textbf{(2)} and \textbf{(3)}. 
\end{enumerate}
\end{minipage} 
}
\caption{Summary of the standard SL QNS protocol for single-axis (dephasing) noise.} 
\label{Protocol1}
\end{table}

\subsection{SPAM errors in spin-locking QNS}
\label{sec:errors} 

Existing QNS protocols are designed under the idealistic assumption that no SPAM errors are present. However, NISQ devices show a high rate of SPAM errors,  
which can significantly affect the estimation accuracy. To understand how these errors enter a SL QNS protocol, we begin from the standard single-axis setting we have just discussed. 

\subsubsection{Modeling state preparation errors}
\label{sub:sp}

Whenever a quantum system is experimentally prepared in an initial state, some degree of error is inevitable. Over-rotations, leakage to states outside the qubit subspace, incomplete relaxation to the ground state, and imperfect projections can all lead to a faulty initial state \cite{PhysRevA.100.032325,Mattias}. While the specific sources of state-preparation (SP) errors are platform-dependent, a faulty initialization may be modeled as a quantum operation mapping the ideal (typically pure) initial state $\rho(0)$ into the corresponding (generally) mixed state $\varrho (0)$ that is actually prepared. 

Throughout this work, we are interested in initial pure states which, in the absence of SP errors, are of the form $\rho_{\ket{u_\pm}}(0)\equiv \ket{u_\pm}\bra{u_\pm}$. If $\varrho_{\ket{u_\pm}}(0)$ denotes the corresponding faulty initial state, a simple model is obtained by letting 
\begin{align}
\label{statepreperr}
    {\varrho}_{\ket{u_\pm}}(0) & \equiv  \tfrac{1}{2} \Big[(1+ \alpha_{SP}) \ket{u_\pm}\bra{u_\pm} + (1-\alpha_{SP}) \ket{u_\mp}\bra{u_\mp} \notag \\
    & + c_u \ket{u_\pm} \bra{u_\mp} + c_u^* \ket{u_\mp} \bra{u_\pm} \Big],
\end{align}
with the SP error-parameter $ \alpha_{SP}$ {\em independent of the direction} $u$ and $|\alpha_{SP}|, |\text{Re}(c_u)|, |\text{Im}(c_u)| \leq 1$. In this way, the state preparation fidelity is 
$$F[ \varrho_{\ket{u_\pm}} (0), \rho_{\ket{u_\pm}}(0)]=\tfrac{1}{2} ({1+\alpha_{SP}}),$$ with $\alpha_{SP}=1$ corresponding to error-free preparation.

\subsubsection{Modeling measurement errors}

In a realistic setting, projective measurements of a state onto a particular basis are likewise not perfect due to systematic errors in distinguishing measurement outcomes or imperfections in the measurement apparatus. As a result, the measurement cannot be simply described by projection operators, and we need to resort to the general formalism of Positive Operator-Valued Measures (POVMs). 

In our analysis, we consider a representation of measurement errors that is in line with current understanding for state-of-the art qubits \cite{PhysRevA.100.052315, PhysRevA.103.042605,Geller_2020,Mattias} and  general enough to capture two leading sources of measurement error: the occurrence of ``false'' readouts; and the ``asymmetry'' in the false readouts. The former are due to overlapping support of outcome probability distributions corresponding to $\ket{z_+}$ and $\ket{z_-}$, which results in readout assignment errors; the latter are the result of qubit decay from $\ket{z_+}$ to  $\ket{z_-}$ during the measurement, making false readouts in $\ket{z_-}$ more likely.

To represent faulty measurements in the $z$ basis, we use the two-outcome POVM  \cite{PhysRevA.103.042605}
\begin{align}
\label{povm01}
    \Pi_{z_+} = \frac{\alpha_M}{2} \sigma_z + \frac{\delta+1}{2} \mathbb{1}, \quad 
    \Pi_{z_-} = \mathbb{1} -  \Pi_{z_+},
 \end{align}
where $0\leq \alpha_M$, $\delta \leq 1$. The parameter $\alpha_M$ relates to the rate of false readouts, whereas $\delta$ relates to the asymmetry in the false readout rate  between the two basis states. When $\alpha_M=1$ and $\delta=0$, the POVM is an ideal projective measurement.  For devices where readout asymmetry is negligible, the model may be simplified to involve only the $\alpha_M$ parameter.

\section{SPAM-robust spin-locking QNS \\
under dephasing noise}
\label{sec:srqnsdephasing}

SPAM errors affect the observable expectation values that are used to infer the spectra in Eqs.\,(\ref{classtheory}) and (\ref{quanttheory}). By computing these expectations in the presence of SPAM errors, we can determine how they affect the estimation. In what follows, we shall use a hat notation to distinguish between the experimentally estimated quantities with SPAM error $\widehat{\braket{\sigma_x(T)}}_{x_\pm}^\Omega$ and the theoretical SPAM-free expectation values $\braket{\sigma_x(T)}_{x_\pm}^\Omega$. The notation will be shortened to
\begin{align*}
\widehat{\braket{\sigma_u(T)}}_{u_\pm}^\Omega&\equiv\widehat{\braket{\sigma_u(T)}}_{\varrho_{\ket{u_\pm}}(0)}^\Omega , \\
    \braket{\sigma_u(T)}_{u_\pm}^\Omega&\equiv\braket{\sigma_u(T)}_{\rho_{\ket{u_\pm}}(0)}^\Omega.
\end{align*}
If measurements are perfect ($\alpha_M=1$, $\delta=0$) and only the SP is faulty, we can determine the expectation values by substituting the faulty initial states $\varrho_{\ket{x_\pm}}(0)$ from Eq.\,\eqref{statepreperr} with $\rho(0)=\ket{\pm x}\bra{\pm x}$ into Eq.\,\eqref{eq::ExpectX}, which yields
\begin{align}\notag
    \widehat{\braket{\sigma_x(T)}}_{x_\pm}^\Omega &= [\varrho_{\ket{x_\pm}}(T)]_{x_+x_+}-[\varrho_{\ket{x_\pm}}(T)]_{x_-x_-} \\
    \label{eqn::expsperr}
&=\braket{\sigma_x(T)}_{x_\pm}^\Omega\mp (1-\alpha_{SP}) e^{-S^+(\Omega)T}.
    \end{align}
Thus, only the amplitude parameter $\alpha_{SP}$, and not the coherence $c_u$, enter the relevant expectation value, and contribute a time-dependent additive bias. It is worth noting that this is valid for the standard case where system-bath couplings are mediated by rank-2 Pauli matrices, as we consider here. If ``rank-1'' (``biased'') couplings would be present \cite{PhysRevA.95.022121, PhysRevB.101.155412}, it would be necessary to also consider unwanted coherences ($c_u$) in modeling faulty SP according to Eq.\,\eqref{statepreperr}.

Similarly, to understand the effect of measurement errors alone, we first assume perfect SP ($\alpha_{SP}=1$) and write the expectation values in terms of measurement outcome probabilities, that is, 
$$\widehat{\braket{\sigma_x(T)}}_{\rho(0)}^\Omega=\mathbb{P}(x_+, \rho(T))-\mathbb{P}(x_-, \rho(T)),$$ 
where $\mathbb{P}(x_\pm, \rho(T))$ is the probability of measuring the qubit in state $\ket{x_\pm}$ given pre-measurement state $\rho(T)$. These probabilities follow by applying a Hadamard gate $H$ to the pre-measurement state and evaluating the POVM in \erf{povm01},
\begin{align}
    \mathbb{P}(x_\pm, \rho(T)) &= \mathrm{Tr}[\Pi_{z_\pm}H\rho(T)H]\notag\\
    \label{eq::measprob}
    &= \frac{1}{2}\Big[ (1\pm\delta) 
    \pm
    \alpha_M \mathrm{Tr}[\sigma_x\rho(T)]\Big].
\end{align} 
Since single-qubit gates can be executed with high fidelity \cite{Yang2019, SHUKLA2020126387}, we neglect errors associated with the Hadamard gate. Identifying  $\text{Tr}[\sigma_x\rho(T)]=\braket{\sigma_x(T)}_{\rho(0)}^\Omega$ as the SPAM-free expectation value, the estimate of the expectation value in the presence of measurement errors is then 
\begin{align}
\label{eq::sxest}
    \widehat{\braket{\sigma_x(T)}}_{\rho(0)}^\Omega &= \alpha_M\braket{\sigma_x(T)}_{\rho(0)}^\Omega + \delta.
\end{align}
We see that $\alpha_M$ adds a multiplicative bias, while $\delta$ introduces an additive bias. 

In the context of SL QNS, the estimate in the presence of both SP and measurement errors can be derived by evaluating the faulty measurement probabilities in Eq.\,\eqref{eq::measprob} with a pre-measurement state $\varrho_{\ket{x_\pm}}(T)$ that evolves from a faulty initial state $\varrho_{\ket{x_\pm}}(0)$. Accordingly, the estimate with the \emph{combined} effect of SPAM errors is
\begin{align*}
    \widehat{\braket{\sigma_x(T)}}_{x_\pm}^\Omega =\mathbb{P}(x_+,\varrho_{\ket{x_\pm}}(T))-\mathbb{P}(x_-,\varrho_{\ket{x_\pm}}(T)),
\end{align*}
whereby it follows that 
\begin{align}
\label{eq::combspam}
    \widehat{\braket{\sigma_x(T)}}_{x_\pm}^\Omega = 
    \alpha_M \Big[\braket{\sigma_x(T)}_{x_\pm}^\Omega \mp (1-\alpha_{SP}) e^{-S^+(\Omega)T}\Big] + \delta .
\end{align}
Having modeled the effect of SPAM errors on the relevant qubit dynamics, we are in a position to seek a modification of the QNS protocol where their effect is removed.

\subsection{Determining SPAM-free classical spectra}

By substituting the SPAM-corrupted expectation values into  Eq.\,\eqref{classtheory}, we obtain a SPAM-corrupted estimate of the classical spectrum,
\begin{align}
    \hat{S}^+(\Omega) =&\, \frac{1}{T}\ln\left[\frac{2}{\widehat{\braket{\sigma_x(T)}}_{x_+}^\Omega-\widehat{\braket{\sigma_x(T)}}_{x_-}^\Omega}\right].\notag
\end{align}
Using Eq.\,\eqref{eq::combspam}, we can relate the right hand-side of this expression to the {\em true} classical spectrum $S^+(\omega)$,
\begin{align}
 \ln\left[\frac{2}{\widehat{\braket{\sigma_x(T)}}_{x_+}^\Omega-\widehat{\braket{\sigma_x(T)}}_{x_-}^\Omega}\right]
    = S^+(\Omega)T - \ln [\alpha], 
    \label{eq::classlinear}
\end{align}
where we have defined a combined SPAM-error parameter $\alpha\equiv\alpha_{SP}\alpha_M$. The fact that SPAM errors contribute a constant, additive term to the estimated classical spectrum is in line with the simple intuition that they enter the qubit dynamics at the start and end of the protocol. The contribution of the classical noise spectrum, in contrast, scales linearly with the total time $T$. We can exploit this relationship to obtain a SPAM-free estimate of the classical spectrum through linear regression in time: The SPAM-free classical spectrum is the resulting slope, while the SPAM parameter $\alpha$ is the intercept.

\subsection{Determining SPAM-free quantum spectra}

By combining the expectation values from Eq.\,\eqref{eq::combspam}, we can create an estimator that relates the ideal quantum and classical spectra as follows: 
\begin{align}
\label{eqn::quantnonlinear}
    \frac{ \widehat{ \braket{\sigma_x(T) }}_{x_+}^\Omega +
    \widehat{\braket{\sigma_x(T)}}_{x_-}^\Omega}{2}
    =\alpha_M \frac{S^-(\Omega)}{S^+(\Omega)} ( 1-e^{-S^+ (\Omega) T } )+\delta.
\end{align}
Two observations can be made. First, the quantum spectrum is coupled to the SPAM parameter $\alpha_M$, implying that stronger measurement errors (smaller $\alpha_M$) effectively {\em reduce} the contribution of the quantum spectrum to the qubit dynamics. Second, as also expected from generalized fluctuation-dissipation arguments \cite{Sarracino,ClerkRMP}, the quantum spectrum is not independent of the classical one: since Eq.\,\eqref{eqn::quantnonlinear} depends on both the classical and quantum spectra, we cannot isolate the quantum spectrum to form an estimator that depends on experimentally measurable quantities alone. Thus, we cannot solve for the quantum spectrum independently through linear regression, as it was possible for the classical spectrum. 

We can make progress, however, by using Eqs.\,\eqref{eq::classlinear}-\eqref{eqn::quantnonlinear} to form a system of equations containing  the unknown parameters $\alpha$, $\alpha_M$, $\delta$, $S^+(\Omega)$, and $S^-(\Omega)$. By fitting the experimental data to these equations, it is possible to estimate {\em some} of these parameters. Still, $\alpha_M$ and $S^-(\Omega)$ are coupled in a way that makes it impossible for them to be estimated independently in the control setting we are interested in. This limitation cannot be overcome by allowing for different qubit initializations or measurements, nor by exploiting steady-state expectations, as we further discuss in Appendix \ref{app:sp}. In order to have a SPAM-robust estimate of the quantum spectrum, either further assumptions about the structure of SPAM contributions are needed, or an independent protocol that can separately pre-characterize SP and measurement error rates \cite{Laflamme2021,Wei2023,Mattias}.

\subsubsection{Exact estimation via non-linear regression}

Here, we shall proceed under the extra assumption that SP errors are small enough to be neglected in comparison to readout errors, which is the case in many qubits of interest \cite{PRXQuantum.2.040326, PhysRevA.103.042605,PhysRevA.100.052315}. If the SP infidelity is expressed as $1-{F}[\varrho_{\ket{u_{\pm}}}(0), \rho_{\ket{u_{\pm}}}(0)]\equiv \epsilon_{SP}$, then it is easy to see that $\alpha_{SP} = 1-2\epsilon_{SP}$. Accordingly, we can let $\alpha=(1-2\epsilon_{SP})\alpha_M \approx \alpha_M$, with accuracy directly related to $\epsilon_{SP}$. In this way, it becomes possible to estimate $\alpha_M$, $\delta$, $S^+(\Omega)$, and $S^-(\Omega)$ via a non-linear regression. We can combine the results into a SPAM-robust QNS protocol for single-axis noise as summarized in Table.\,\ref{Protocol2}.

\noindent
\begin{table}
\setlength{\fboxrule}{0.8pt}
\fbox{
\begin{minipage}[t][7.4cm]{0.46\textwidth} 
\raggedright 
        \mbox{\bf{Protocol 2: SPAM-robust single-axis spin-locking QNS} }\\
        \rule{8cm}{0.5pt}
        \noindent \mbox{\textit{Input.} Driving amplitude $\Omega$, a set of times $\{T_j\} $.}\\
        \vspace*{1mm}
        \noindent \mbox{\textit{Output.} Noise spectra at frequency $\omega=\Omega$ and SPAM parameters.} \\     
        \vspace*{1mm}   
        \noindent \mbox{\textbf{The protocol:} \hspace*{2cm}     }          
\begin{enumerate}
    \item[{\bf (1)}] Prepare the initial state $\ket{x_+}\bra{x_+}$, apply a resonant drive along $x$ with amplitude $\Omega$, let the qubit evolve for time $T_j$ such that $|\Omega| T_j\gg 1$, and measure $\sigma_x$.
    \item[{\bf (2)}] Repeat \textbf{(1)} many times to estimate $\widehat{\braket{\sigma_x(T_j)}}_{x_+}^\Omega$.
    \item[{\bf (3)}] Repeat \textbf{(1)} and \textbf{(2)} with the qubit prepared in $\ket{x_-}\bra{x_-}$ to estimate $\widehat{\braket{\sigma_x(T_j)}}_{x_-}^\Omega$.
    \item[{\bf (4)}] Repeat steps \textbf{(1)}-\textbf{(3)}
    for all $T_j\in \{T_j\}$ to create the sets of expectation values $\{\widehat{\braket{\sigma_x(T_j)}}_{x_+}^\Omega\}$ and $\{\widehat{\braket{\sigma_x(T_j)}}_{x_-}^\Omega\}$.
    \item[{\bf (5)}] Fit the data to the Eqs.\,\eqref{eq::classlinear}-\eqref{eqn::quantnonlinear} to estimate $S^+(\Omega)$, $S^-(\Omega)$, $\alpha_M$ and $\delta$, under the approximation $\alpha\approx\alpha_M$.
    \end{enumerate}
\end{minipage}
}
\caption{Summary of the proposed SPAM-robust SL QNS protocol for single-axis (dephasing) noise.} 
\label{Protocol2}
\end{table}

\begin{figure*}
    \centering
    \includegraphics[scale=0.8]{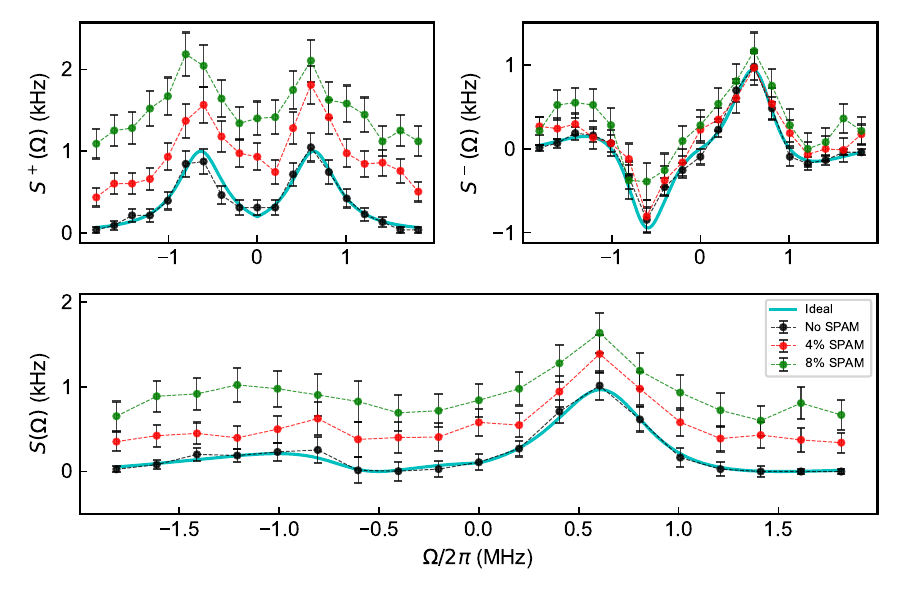}
    \vspace*{-6mm}
    \caption{(Color online) {\bf Effect of SPAM errors on spectral reconstruction in a simulated single-axis QNS experiment.} Performance of standard SL QNS protocol (Table\,\ref{Protocol1}) in the presence of varying SPAM error rates and dephasing noise as in Eqs.\,\eqref{eq:dephsp}-\eqref{eq:lor}, with parameters $\omega_0=4.0/2\pi \text{ MHz}$ and $t_c=0.5\text{ $\mu$s}$. Top panels: Reconstructed classical (left) vs. quantum (right) spectra. Bottom panel: Complete spectra. In all plots, the black lines show the ideal case with no SPAM-errors and the injected target noise spectrum is represented by the cyan line.  The injected SPAM-error parameters are $\alpha_{SP}=0.99$, $\alpha_M=0.97$, $\delta=0.01$ (red); $\alpha_{SP}=0.98$, $\alpha_M=0.94$, $\delta=0.02$ (green). The relevant observable expectation values are evaluated through averages over a 1,000 shots, with error bars representing a $95\%$ confidence-interval.}
\label{fig1}
\end{figure*}

\subsubsection{Approximate estimation via linearized regression}
\label{smallspt}

If, in the process of estimating the classical spectrum from Eq.\,\eqref{eq::classlinear}, it is observed that the experimentally determined estimates $\{\hat S^+(\Omega)\}$ obey the condition $\max_{T_j} \{\hat{S}^+(\Omega)T_j\}\ll 1$, a simplified estimation procedure is possible. Specifically, we can make use of the approximation
\begin{align}
\label{approx}
    \frac{S^+(\Omega)T}{1-e^{- S^+(\Omega)T}} \approx 1 + O(S^+(\Omega)T),
\end{align}
meaning that \erf{eqn::quantnonlinear} can be simplified to
\begin{align}
\label{eq::quantlinear}
    \frac{\widehat{\braket{\sigma_x(T)}}_{x_+}^\Omega + \widehat{\braket{\sigma_x(T)}}_{x_-}^\Omega}{2}&\approx \alpha_M S^-(\Omega)T + \delta .
\end{align}
By treating the above approximate equality as an exact one, the combination of Eqs.\,\eqref{eq::classlinear} and \eqref{eq::quantlinear} allows us to estimate the classical spectrum, the (scaled) quantum spectrum and the SPAM parameters $\alpha,\delta$ through a linear regression.

In addition to simplifying the fitting problem, it is worth noting that such an approximate estimation procedures allows us to pinpoint the effect of the total evolution time $T$ in the SPAM-induced errors. We find that the SPAM-induced additive biases to the estimated spectra scale inversely with $T$, 
\begin{align}
    \label{eq::wings1}
    \hat S^+(\Omega)& = S^+(\Omega)-\tfrac{1}{T}\ln[\alpha_M],\\
    \label{eq::wings2}
    \hat S^-(\Omega)& = \alpha_M S^-(\Omega) - \tfrac{1}{T}{\delta},
\end{align}
where, in this regime, the estimate of the quantum spectrum simplifies as $S^-(\Omega)\equiv\tfrac{1}{2}({\widehat{\braket{\sigma_x(T)}}_{x_+}^\Omega + \widehat{\braket{\sigma_x(T)}}_{x_-}^\Omega})$. Thus, the additive biases vanish in the steady state. However, to estimate the classical and the quantum spectra independently, the protocol operates in the ``long-but-finite-time'' regime, where it is necessary to factor in the competing effects of the evolution time $T$ and the SPAM-error rates to the bias. This will be relevant in interpreting the experimental results of Sec.\,\ref{sec::multiqnsprotocol}.

\subsection{Dephasing protocol validation: Numerical \\
simulation results}

In order to quantitatively assess the impact of SPAM errors, and to test our SPAM-robust protocol in the simpler single-axis setting, we performed numerically simulated QNS experiments. To simulate a non-classical dephasing bath with a colored continuous spectrum, we consider a toy model where the bath comprises a single qubit, and interacts with the system qubit via a fluctuating coupling operator that involves two non-commuting directions, namely, 
\begin{align}
\label{eq::couplingHsim}
    H(t)= \tfrac{1}{2} \sigma_z\otimes \big[ \beta(t) \tau_x + {\beta(t+\gamma)} \tau_y\big].
\end{align}    
Here, $\tau_i$ are Pauli operators acting on the bath qubit, $\gamma$ is a lag time, and $\beta(t)$ is assumed to be a classical zero-mean stationary Gaussian process (see Appendix\,\ref{quantnoisesim} for additional detail, including a more general Hamiltonian). The quantum and classical spectra generated by $H(t)$ depend on the spectrum of $\beta(t)$, which we denote by $\Tilde{S}(\omega)$. Using the definitions in Eqs.\,\eqref{eq::spec}, \eqref{eq::specquant} and \eqref{eq::specclass}, we obtain the following spectra: 
\begin{align}
    S^+(\omega)&=\Tilde{S}(\omega),\quad
    S^-(\omega)=\Tilde{S}(\omega)\sin(\gamma \omega), \notag \\
    &S(\omega)=\tfrac{1}{2}\Tilde{S}(\omega)[1+\sin(\gamma \omega)].
    \label{eq:dephsp}
\end{align}
Note that a lag time $\gamma>0$ between the fluctuations coupled to $\tau_x$ and $\tau_y$ is needed to ensure a non-vanishing quantum spectrum. In our simulations, we choose the classical process $\beta(t)$ to have a Lorentzian spectrum, of the form
\begin{align}
    \Tilde{S}(\omega) = \frac{1}{1+t_c^2(|\omega|-\omega_0)^2}, 
\label{eq:lor}
\end{align}

\begin{figure*}
    \centering
    \includegraphics[scale=0.8]{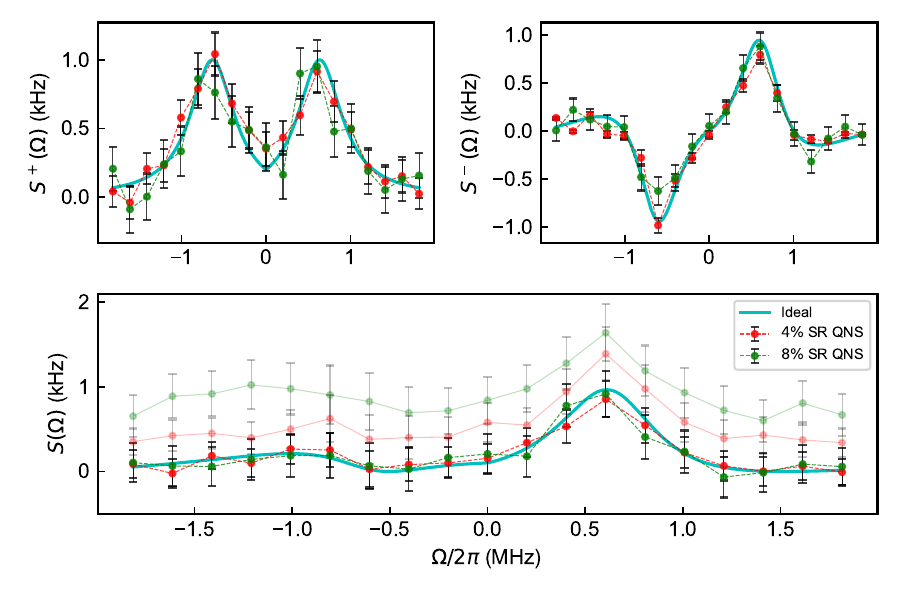}
    \vspace*{-6mm}
    \caption{(Color online) {\bf Performance of SPAM-robust single-axis QNS.} The plot shows a performance comparison between the standard SL QNS protocol (described in Table\,\ref{Protocol1}; translucent lines) and the SPAM-robust SL QNS protocol (labelled SR QNS, described in Table\,\ref{Protocol2}; dotted lines) under varying rates of SPAM errors and dephasing noise as in Fig.\,\ref{fig1}. Top panels: Reconstructed classical (left) vs. quantum (right) spectra. Bottom panel: Complete dephasing spectra. While the product $\alpha_{SP}\alpha_M$ is the same as in Fig.\,\ref{fig1}, here $\alpha_{SP}=1$ and all the contribution comes from measurement errors, with $\alpha_M=0.96$, $\delta=0.01$ (red), and $\alpha_M=0.92$, $\delta=0.02$ (green). In all cases, the ideal noise spectrum is represented by the solid cyan line. The relevant observable expectation values are evaluated through averages over a 1,000 shots, with the error bars representing a $95\%$ confidence-interval; a set of 15 different times $\{T_j\}_{j=1}^{15}$ are used in the SR QNS implementation.}
\label{fig2}
\end{figure*}

\begin{figure*}
    \centering
    \includegraphics[scale=0.75]{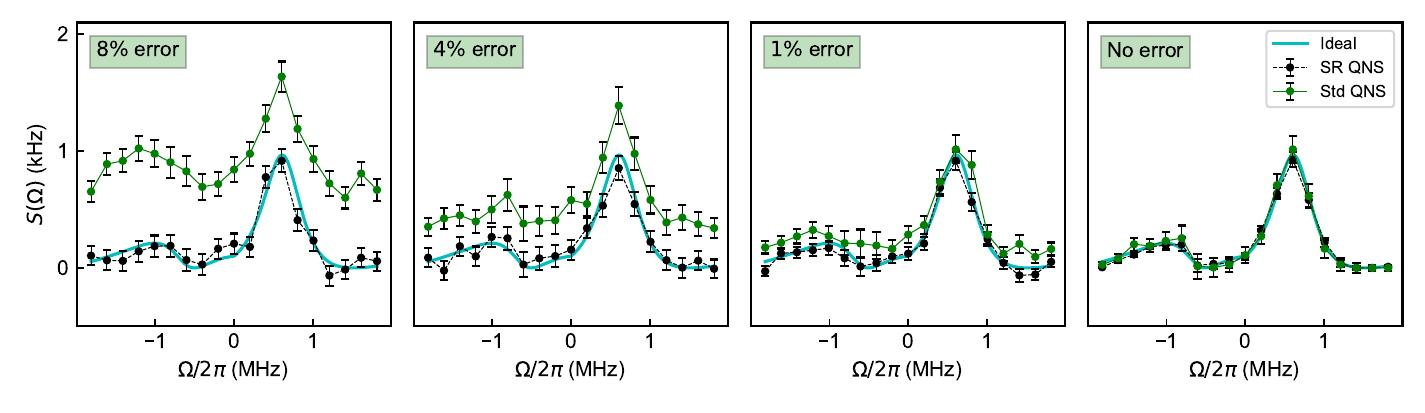}
    \vspace*{-6mm}
    \caption{(Color online) {\bf SPAM-robust single-axis QNS performance for different error rates.} The green lines show the results from standard SL QNS, while the black lines show the results SPAM-robust QNS. From left to right, the SPAM error rates are $8\%$ ($\alpha_M=0.92, \delta= 0.02$), $4\%$ ($\alpha_M=0.96, \delta =0.01$), $1\%$ ($\alpha_M=0.99, \delta =0.005$), and no SPAM errors. Dephasing noise parameters are as in previous figures.}
    \label{fig3}
\end{figure*}

Figure\,\ref{fig1} shows how the spectra reconstructed from a standard SL QNS protocol are affected by SPAM errors of variable strength. In particular, the classical spectrum (top left) is uniformly offset from the ideal spectrum by the presence of $\alpha=\alpha_{SP}\alpha_M$, whereas the quantum spectrum (top right) shows {\em both} an offset by $\delta$ and a re-scaling by $\alpha_M$, respectively. In Fig.\,\ref{fig2}, the same values of $\alpha$ are introduced into the simulation, however now the SPAM-robust QNS protocol is numerically implemented to mitigate the effects of these errors. The simulation allows us to make the $\max_{T_j}\{\hat S^+(\Omega)T_j\}\ll 1$ approximation, so we use the linearized regression procedure described in Sec.\,\ref{smallspt} to estimate the SPAM-free spectra. The spectral reconstructions are now found to be in excellent agreement for both the ideal classical and quantum spectra. Furthermore, the protocol allows us to estimate the SPAM parameters to a high precision, as reported in the caption.

Taken together, Figs.\,\ref{fig1} and \ref{fig2} are effective at demonstrating the advantages of the SPAM-robust QNS, however they also show that the overhead of the SPAM-robust QNS need not be warranted if the SPAM error is below a certain level. In Fig.\,\ref{fig3}, we report reconstructions of the complete spectra for progressively smaller SPAM error rates, showing how some advantage of SPAM-robust QNS is retained even at the $1\%$ error rate. Ultimately, the usefulness of a SPAM-robust protocol will depend on the competition between the SPAM-errors and the evolution time, with the advantage of SPAM-robust QNS diminishing with a decreasing rate of SPAM-errors (recall Eqs.\,\eqref{eq::wings1}-\eqref{eq::wings2}).

\section{Spin-locking QNS under multi-axis noise}
\label{sec::slqnsmultiaxis}

\subsection{Driven qubit dynamics}

In order to design a generalized SL QNS protocol that can operate under multi-axis qubit noise, assuming a weak-coupling regime, it is necessary to consider the TCL ME of Eq.\,\eqref{eq::TCL2ndOrder}, with a controlled open-system Hamiltonian of the general form given in Eq.\,\eqref{eq::HTogglingMulti}. Let us introduce matrix elements of the toggling-frame density operator in the qubit eigenbases defined in Eq.\,\eqref{eigs}, that is, 
$$\rho_{u_j u_{j'}} (t) \equiv \langle u_j |\rho(t) |u_{j'}\rangle, \quad u=x,y,z, \;j,j'=\pm.$$
By transforming to the frequency domain, we can show that, under a secular approximation that removes rapidly oscillating terms, the ME can be simplified as follows (see Appendix \ref{appendix::tclmema} for a detailed derivation and additional discussion):
\begin{widetext}
\begin{eqnarray}
\dot{\rho}_{u_j u_{j'}} (t) &=&\!-\frac{1}{2\pi} \!\!\sum_{\substack{\alpha,\beta,\beta'\\=-1,0,1}}\int_{-\infty}^\infty\!\!d\omega\,e^{i\omega t}y_{\alpha\beta}(t)\!\!\int_0^t\!\!ds\,e^{-i\omega s}y_{-\alpha\beta'}(s) \label{eq::SecularApprox}\\ 
&\times & \bigg\{ \sum_{\ell=\pm 1}\; 
\Big[S_{\alpha,-\alpha}(-\omega+\alpha \omega_q)\bra{u_\ell}\sigma_{\beta'}\sigma_{\beta}\ket{u_{j'}}\rho_{u_ju_\ell}(t)
+S_{-\alpha,\alpha}(\omega-\alpha \omega_q)\bra{u_j}\sigma_{\beta}\sigma_{\beta'}\ket{u_\ell}\rho_{u_\ell u_{j'}}(t)\Big] \notag\\
\notag 
&- &\!\!\!\!\sum_{\ell,\ell'=\pm 1}
\Big[S_{\alpha,-\alpha}(-\omega+\alpha \omega_q)\bra{u_\ell}\sigma_{\beta'}\ket{u_{j'}}\bra{u_j}\sigma_{\beta}\ket{u_{\ell'}}
+S_{-\alpha,\alpha}(\omega-\alpha \omega_q)\bra{u_\ell}\sigma_{\beta}\ket{u_{j'}}\bra{u_j}\sigma_{\beta'}\ket{u_{\ell'}}\Big]\rho_{u_{\ell'}u_\ell}(t) \bigg \}. \notag
\end{eqnarray}
\end{widetext}
This provides the most general expression that can be obtained without specifying the precise form of the applied control Hamiltonian, hence the switching functions $y_{\alpha\beta}(t)$. Note that, while the dephasing spectrum ($\alpha=0$) enters the dynamics via $S_{0,0}(\omega)$, the transverse spectra are shifted by the qubit frequency as $S_{\alpha,-\alpha}(-\omega+\alpha\omega_q)$ and $S_{-\alpha,\alpha}(\omega-\alpha\omega_q)$, respectively. Physically, this reflects the fact that transverse terms with frequencies close to the transition frequency $\omega_q$ contribute to processes of qubit excitation (absorption) and de-excitation (emission), while frequencies far off-resonance are not relevant to the dynamics under the stated assumptions.

\subsection{Spin-locking QNS protocol}

We first show how, starting from Eq.\,\eqref{eq::SecularApprox}, we may build a multi-axis SL QNS protocol in the absence of SPAM-errors.  
Although the above-mentioned secular approximation removes terms that are rapidly oscillating in frequency, solving the set of coupled DEs for arbitrary time-dependent control is not possible. Fortunately, these expressions simplify when we restrict the control to a constant drive along a fixed axis, i.e., $H_\text{ctrl}=\tfrac{1}{2}\Omega\sigma_z$ or $H_\text{ctrl}=\tfrac{1}{2}\Omega\sigma_x$. In this case, as we show in Appendix\,\ref{appendix::tclmema}, the coefficients in Eq.\,(\ref{eq::SecularApprox}) are approximately time-independent in the limit $|\Omega| t\gg 1$. Thus, a corresponding set of linear first-order DEs can be derived. Specifically, for a constant drive along $\sigma_z$, this set is 
\begin{eqnarray*}
\dot\rho_{z_+z_+}(t) &= & -2S_{-1,1}(-\Omega-\omega_q)\rho_{z_+z_+}(t)\\
&+&2S_{1,-1}(\Omega+\omega_q)\rho_{z_-z_-}(t) ,\\
\dot\rho_{z_+z_-}(t) &= &-\Big[S_{-1,1}(-\omega_q-\Omega)+S_{1,-1}(\omega_q+\Omega)\\
&+&2S_{0,0}(0)\Big] \rho_{z_+z_-}(t).
\end{eqnarray*}
Driving along $\sigma_x$ produces instead 
\begin{eqnarray*}
\dot\rho_{x_+x_+}(t)-\dot\rho_{x_-x_-}(t)&\approx &
- A(\Omega)\left[ \rho_{x_+x_+}(t)-\rho_{x_-x_-}(t)\right]\\ 
&+&B(\Omega),
\end{eqnarray*}
where we have defined the following combinations of spectra: 
\begin{widetext}
\begin{align}
\label{Adef}
    A(\Omega)&\equiv S_{0,0}(\Omega)+S_{0,0}(-\Omega)+\frac{1}{2} \Big[S_{1,-1}(\Omega+\omega_q)+S_{-1,1}(-\Omega-\omega_q)+ S_{1,-1}(-\Omega+\omega_q)+ S_{-1,1}(\Omega-\omega_q) \Big], \\
\label{Bdef}
    B(\Omega)&\equiv S_{0,0}(\Omega)-S_{0,0}(-\Omega)+\frac{1}{2}\Big[ S_{1,-1}(\Omega+\omega_q)-S_{-1,1}(-\Omega-\omega_q)+S_{1,-1}(-\Omega+\omega_q)- S_{-1,1}(\Omega-\omega_q) \Big].
\end{align}
\end{widetext}
By making the appropriate frame transformation, we can write the rotating-frame DEs under a constant $\sigma_z$ drive,
\begin{eqnarray*}
\dot\rho_{\text{rot, }z_+z_+}(t) &=&-2S_{-1,1}(-\Omega-\omega_q)\rho_{\text{rot, }z_+z_+}(t)\\
&+&2S_{1,-1}(\Omega+\omega_q)\rho_{\text{rot, }z_-z_-}(t), \\
\end{eqnarray*}
\begin{eqnarray*}
\dot\rho_{\text{rot, }z_+z_-}(t) &=& -\Big[S_{-1,1}(-\omega_q-\Omega)+S_{1,-1}(\omega_q+\Omega)\\
&+&2S_{0,0}(0)\Big] \rho_{\text{rot, }z_+z_-}(t)-i\Omega\rho_{\text{rot, }z_+z_-}(t),
\end{eqnarray*}
and a constant $\sigma_x$ drive, 
\begin{eqnarray*}
\dot\rho_{\text{rot, }x_+x_+}(t)-\dot\rho_{\text{rot, }x_-x_-}(t) &= &
- A(\Omega) \big[\rho_{\text{rot, }x_+x_+}(t)  \\ 
&-& \rho_{\text{rot, }x_-x_-}(t)\big]  + B(\Omega).
\end{eqnarray*}
As described in Sec.\,\ref{sec::background}, the dynamics of the coherence elements are not frame-invariant for arbitrary times. By ensuring, however, that the evolution time obeys a ``frame-alignment condition'' $T^{(n)}\equiv 2\pi n/|\Omega|$, $n\in \mathbb{N}$, the full qubit dynamics in the (physical) rotating frame can be (stroboscopically) accessed from the ones computed in the toggling frame.

In the toggling frame, the above DEs can be written in terms of observable expectation values by relating 
\begin{eqnarray*}
&&\rho_{z_+z_-}(t)+\rho_{z_-z_+}(t)\equiv\braket{\sigma_x(t)}^\Omega_{\rho(0)}, \\
&&\rho_{z_+z_+}(t)-\rho_{z_-z_-}(t)\equiv \braket{\sigma_z(t)}^\Omega_{\rho(0)},
\end{eqnarray*}
and subsequently solved for the expectation values. By now changing the superscript $\Omega \mapsto \Omega_u$ to explicitly denote the direction of the applied drive, and by using the equations for the constant $z$-drive, we find 
\begin{align}
\braket{\sigma_z(t)}^{\Omega_z}_{\rho(0)}&=\frac{e^{-2[S_{-1,1}(-\Omega-\omega_q)+S_{1 ,-1}(\Omega+\omega_q)]t}}{S_{-1,1}(-\Omega-\omega_q)+S_{1, -1}(\Omega+\omega_q)}\notag\\
    &\times\Big\{2[\rho_{z_+z_+}(0)S_{-1,1}(-\Omega-\omega_q)\notag\\
    &-\rho_{z_-z_-}(0)S_{1, -1}(\Omega+\omega_q)]\notag\\
    &-e^{2[S_{-1,1}(-\Omega-\omega_q)+S_{1 ,-1}(\Omega+\omega_q)]t}\notag\\
    \label{eq::szHz}
    &\times[S_{-1,1}(-\Omega-\omega_q)-S_{1 ,-1}(\Omega+\omega_q)]\Big\},
\end{align}
which depends only on the transverse spectra, and
\begin{align}
\braket{\sigma_x(t)}_{\rho(0)}^{\Omega_z}&=\left[\rho_{z_+z_-}(0)+\rho_{z_-z_+}(0)\right]
    \label{eq::togsxsol} \\
    &\times e^{-\left[S_{-1,1}(-\Omega-\omega_q)+S_{1, -1}(\Omega+\omega_q)+2S_{0,0}(0)\right]t}.\notag
\end{align}
This observable permits the dephasing spectrum to be estimated at $S_{0,0}(0)$. Likewise, 
\begin{align}
\label{slqnsdephasing}
    \braket{\sigma_x(t)}^{\Omega_x}_{\rho(0)}=\frac{e^{-A(\Omega)t}[B(\Omega)(e^{A(\Omega)t}-1)+A(\Omega)\braket{\sigma_x(0)}]}{A(\Omega)}, 
\end{align}
which shows how, when the qubit is driven off-axis, the dynamics are dictated by both the transverse and the dephasing spectra. As one may verify, Eq.\,\eqref{eq::ExpectX} can be recovered in the limiting case of dephasing-only noise, where $A(\Omega)=S^+_{0,0}(\Omega)$ and $B(\Omega)=S^-_{0,0}(\Omega)$.

\noindent
\begin{table*}[t]
\setlength{\fboxrule}{0.8pt}
\fbox{
\begin{minipage}[t][17cm]{0.8\textwidth}
\raggedright 
        \mbox{\bf{Protocol 3: Multi-axis spin-locking QNS} }\\
        \rule{14.2cm}{0.5pt}
        \noindent \mbox{\textit{Input.} Driving amplitude $\Omega$.}\\  
        \vspace*{1mm}
        \noindent \mbox{\textit{Output.} Noise spectra at frequency $\Omega$ and SPAM parameters.} \\
        \vspace*{1mm}   
        \noindent \mbox{\textbf{The protocol:} \hspace*{2cm}     }  
\begin{enumerate}
\item[{\bf (1)}] Implement single-axis QNS with $H_\text{ctrl} = \tfrac{1}{2}\Omega\sigma_z$:
\vspace*{-0mm}
\begin{enumerate}
\item[\textbf{a)}] Prepare the initial state $\ket{z_+}\bra{z_+}$, apply a resonant drive along $z$ with amplitude $\Omega$. Let the qubit evolve for time $T$, with $|\Omega| T\gg 1$, and measure $\sigma_z$.
\item[\textbf{b)}] Repeat \textbf{a)} sufficiently many times to estimate $\braket{\sigma_z(T)}^{\Omega_z}_{z_+}$.
\item[\textbf{c)}] Repeat steps \textbf{a)} and \textbf{b)} with the qubit prepared in $\ket{z_-}\bra{z_-}$, to estimate $\braket{\sigma_z(T)}_{z_-}^{\Omega_z}$.
\item[\textbf{d)}] Prepare the initial state $\ket{x_+}\bra{x_+}$, apply a resonant drive along $z$ with amplitude $\Omega$. Let the qubit evolve for time $T^{(n)}$, with $|\Omega| T^{(n)}\gg 1$, $T^{(n)} = 2\pi n/|\Omega|$, for some $n\in {\mathbb N}$, and measure $\sigma_z$. 
\item[\textbf{e)}] Repeat \textbf{d)} sufficiently many times to estimate $\braket{\sigma_x(T^{(n)})}^{\Omega_z}_{x_+}$.
\item[\textbf{f)}] Repeat steps \textbf{d)} and \textbf{e)} with the qubit prepared in $\ket{x_-}\bra{x_-}$, to estimate $\braket{\sigma_x(T^{(n)})}^{\Omega_z}_{x_-}$.
\end{enumerate}

\item[{\bf (2)}] Use the expectation values and Eqs. \eqref{eq::czdrp}-\eqref{eq::qzdrp} to infer the classical spectrum $S_{1,-1}^+(\Omega+\omega_q)$, the quantum spectrum $S_{-1,1}^-(-\Omega-\omega_q)$, and the dephasing spectrum $S^+_{0,0}(0)$.

\item[{\bf (3)}] Implement single-axis QNS with $H_\text{ctrl} = -\tfrac{1}{2}\Omega\sigma_z$:
\begin{enumerate}
\item[\textbf{a)}] Prepare the initial state $\ket{z_+}\bra{z_+}$, apply a resonant drive along $z$ with amplitude $\Omega$. Let the qubit evolve for time $T$ such that $|\Omega| T\gg 1$, and measure $\sigma_z$.
\item[\textbf{b)}] Repeat \textbf{a)} sufficiently many times to estimate $\braket{\sigma_z(T)}^{-\Omega_z}_{z_+}$.
\item[\textbf{c)}] Repeat steps \textbf{a)} and \textbf{b)} with the qubit prepared in $\ket{z_-}\bra{z_-}$, to estimate $\braket{\sigma_z(T)}_{z_-}^{-\Omega_z}$.
\end{enumerate}

\item[{\bf (4)}] Use the expectation values and Eqs. \eqref{eq::czdrm}-\eqref{eq::qzdrm} to infer the classical spectrum $S_{-1,1}^+(\Omega-\omega_q)$ and the quantum spectrum $S_{1,-1}^-(-\Omega+\omega_q)$.

\item[{\bf (5)}] Implement single-axis QNS with $H_\text{ctrl} = \tfrac{1}{2}\Omega\sigma_x$: 
\vspace*{-0mm}
\begin{enumerate}
\item[\textbf{a)}] Prepare the initial state $\ket{x_+}\bra{x_+}$, apply a resonant drive along $x$ with amplitude $\Omega$. Let the qubit evolve for 
time $T$, with $|\Omega| T \gg 1$, and measure $\sigma_x$.
\item[\textbf{b)}] Repeat \textbf{a)} sufficiently many times to estimate $\braket{\sigma_x(T)}^{\Omega_x}_{x_+}$.
\item[\textbf{c)}] Repeat steps \textbf{a)} and \textbf{b)} with the qubit prepared in $\ket{x_-}\bra{x_-}$, to estimate $\braket{\sigma_x(T)}_{x_-}^{\Omega_x}$.
\end{enumerate}

\item[{\bf (6)}]Infer $A(\Omega)$ and $B(\Omega)$ from Eqs.\,\eqref{eq::cxdr}-\eqref{eq::qxdr}. Use Eqs.\,(\ref{Adef}) and (\ref{Bdef}) with the inferred transverse spectra to estimate the classical and quantum dephasing spectra $S^+_{0,0}(\Omega)$ and $S^-_{0,0}(\Omega)$:
\begin{itemize}
    \item $S^+_{0,0}(\Omega)= A(\Omega)-\frac{1}{2}\left[S^+_{1,-1}(\Omega+\omega_q)+S^+_{-1,1}(\Omega-\omega_q)\right]$,
    \item $S^-_{0,0}(\Omega)= B(\Omega)-\frac{1}{2}\left[S^-_{1,-1}(-\Omega+\omega_q)-S^-_{-1,1}(-\Omega-\omega_q)\right]$.
\end{itemize}   
\end{enumerate}
\end{minipage} }
\caption{Summary of the proposed SL QNS protocol for multi-axis noise.}
\label{Protocol3} 
\end{table*}

With the appropriate choice of initial states, a set of independent equations may be derived for the observable expectation values, which can then be inverted to infer the classical and the quantum spectra, given in Eqs.\,\eqref{eq::specquant2}-\eqref{eq::specclass2}.
In fact, it becomes clear that multi-axis QNS in the spherical basis entails, in essence, a repetition of three single-axis QNS protocols. 


First, by letting $H_\text{ctrl}=\tfrac{1}{2}\Omega\sigma_z$, Eqs.\,(\ref{eq::szHz})-(\ref{eq::togsxsol}) with the initial states $\{\ket{z_+},\ket{z_-},\ket{x_+},\ket{x_-}\}$ generate independent solutions that relate the spherical spectra to observable expectation values via linear combinations of the resulting solutions:
\begin{eqnarray}
\label{eq::czdrp}
    S_{1,-1}^+(\Omega+\omega_q)=\frac{1}{T}\ln\left[\frac{2}{\braket{\sigma_z(T)}^{\Omega_z}_{z_+}-\braket{\sigma_z(T)}^{\Omega_z}_{z_-}}\right],
\end{eqnarray}
\vspace*{-5mm}
\begin{eqnarray}    
\label{eq::czdrxp}
   S_{0,0}(0) &=&\frac{1}{2 T^{(n)}}\ln\left[\frac{2}{\braket{\sigma_x(T^{(n)})}^{\Omega_z}_{x_+}-\braket{\sigma_x(T^{(n)})}^{\Omega_z
    }_{x_-}}\right]\notag\\
    &-&\frac{1}{2}S_{1-1}^+(\Omega+\omega_q), 
\end{eqnarray}
\vspace*{-5mm}
\begin{eqnarray}
    S_{-1, 1}^-(-\Omega-\omega_q)&=& \frac{\braket{\sigma_z(T)}_{z_+}^{\Omega_z}+\braket{\sigma_z(T)}_{z_-}^{\Omega_z}}{2}\notag\\
    \label{eq::qzdrp}
    &\times&\frac{S_{1 ,-1}^+(\Omega+\omega_q)}{e^{-2S_{1 ,-1}^+(\Omega+\omega_q)T}- 1}.
\end{eqnarray}

Second, with the opposite $z$-drive, $H_\text{ctrl}=-\tfrac{1}{2}\Omega\sigma_z$ and the initial states $\{\ket{z_+},\ket{z_-}\}$, we can express the remaining classical and quantum transverse spectra in terms of observable expectation values as follows:
\begin{eqnarray}
\label{eq::czdrm}
    S_{-1,1}^+(\Omega-\omega_q)=\frac{1}{T}\ln\!\left[\frac{2}{\braket{\sigma_z(T)}^{-\Omega_z}_{z_+}-\braket{\sigma_z(T)}^{-\Omega_z}_{z_-}}\right]\!,
\end{eqnarray}
\vspace*{-5mm}
\begin{eqnarray}    
    S_{ 1, - 1}^-(-\Omega+\omega_q)&=& \frac{\braket{\sigma_z(T)}_{z_+}^{-\Omega_z}+\braket{\sigma_z(T)}_{z_-}^{-\Omega_z}}{2} \notag\\
    \label{eq::qzdrm}
    &\times&\frac{S_{- 1  1}^+(\Omega-\omega_q)}{1 - e^{-2S_{- 1  1}^+(\Omega-\omega_q)T}}.
\end{eqnarray}

Lastly, with the $x$-drive Hamiltonian $H_\text{ctrl}=\tfrac{1}{2}\Omega\sigma_x$, Eq.\,\eqref{slqnsdephasing} can be solved for the initial states $\{\ket{x_+}, \ket{x_-}\}$, allowing us to relate $A(\Omega)$ and $B(\Omega)$ to the observable expectation values
\begin{align}
\label{eq::cxdr}
    A(\Omega)&=\frac{1}{T}\ln\left[\frac{2}{\braket{\sigma_x(T)}^{\Omega_x}_{x_+}-\braket{\sigma_x(T)}^{\Omega_x}_{x_-}}\right],\\
   \label{eq::qxdr} B(\Omega)&=\frac{\braket{\sigma_x(T)}_{x_+}^{\Omega_x}+\braket{\sigma_x(T)}_{x_-}^{\Omega_x}}{2} \left( \frac{A(\Omega)}{1-e^{-A(\Omega)T}}\right).
\end{align}
Having estimated the transverse spectra, it is possible to infer the classical and quantum dephasing spectra from the definitions of $A(\Omega)$ and $B(\Omega)$ in Eq.\,\eqref{Adef}-\eqref{Bdef}. The complete protocol may be summarized in Table\,\ref{Protocol3}.

\section{SPAM-robust spin-locking QNS \\ under multi-axis noise} 
\label{sec::srqnsmultiaxis}

While SPAM errors enter into the qubit dynamics in a structurally similar way as they did under single-axis dephasing noise, the multi-axis QNS protocol entails the estimation of the expectation values $\widehat{\braket{\sigma_x(t)}}^{\Omega_x}_{x_\pm}$ (under the $x$-drive, as in the dephasing case) and, additionally,  $\widehat{\braket{\sigma_z(t)}}^{\Omega_z}_{z_\pm}$ and $\widehat{\braket{\sigma_x(t)}}^{\Omega_z}_{x_\pm}$ (under the longitudinal $z$-drives). Similarly to Eq.\,\eqref{eq::combspam}, the SPAM-modified observable expectation values after an evolution time $T$ now take the form 
\begin{eqnarray}
\label{eq::szhatma}
\widehat{\braket{\sigma_z(T)}}^{\Omega_z}_{z_\pm}&= & \alpha_M\Big[\braket{\sigma_z(T)}^{\Omega_z}_{z_\pm} \notag\\
&\mp & \left(1\pm\alpha_{SP}\right)e^{-S^+_{\pm1\mp1}(\Omega\pm\omega_q)T}\Big]+\delta,\quad \\
\label{eq::sxhatma1}
\widehat{\braket{\sigma_x(T)}}^{\Omega_{z}}_{x_\pm}&=& \alpha\braket{\sigma_x(T)}^{\Omega_{z}}_{x_\pm}+\delta, 
\end{eqnarray}
\begin{eqnarray}
\widehat{\braket{\sigma_x(T)}}^{\Omega_x}_{x_\pm}&=& \alpha_M\Big[\braket{\sigma_x(T)}^{\Omega_x}_{x_\pm}\notag\\
\label{eq::sxhatma2}
&\mp & \left(1\pm\alpha_{SP}\right)e^{-A(\Omega)T}\Big]+\delta.
\end{eqnarray}

\subsection{Determining SPAM-free classical spectra}

The estimates of the classical spectra can be written in terms of the estimates of the observable expectation values and then related to the ideal, SPAM-free transverse and dephasing spectra. By substituting the appropriate observable expectation values from Eqs.\,\eqref{eq::szhatma}-\eqref{eq::sxhatma2} into Eqs.\,\eqref{eq::czdrp}-\eqref{eq::qxdr}, we can relate the SPAM-modified estimates of the classical spectra to the corresponding ideal quantities. Specifically, we have 
\begin{eqnarray}    
 &&    \ln\left[\frac{2}{\braket{\widehat{\sigma_x(T^{(n)})}}^{\Omega_z}_{x_+}-\widehat{\braket{\sigma_x(T^{(n)})}}^{\Omega_z}_{x_-}}\right]=  2 
    S_{0,0}(0)T^{(n)} \notag\\
&&  \hspace*{0mm}  
 + S_{-1,1}(-\Omega-\omega_q)+  \,S_{1,-1}^+(\Omega+ \omega_q)T^{(n)}-\ln[\alpha], 
      \label{eq::dephspamfirst}
\end{eqnarray}
where the condition $\{T^{(n)}_j=2\pi n_j/|\Omega|\}$ is needed in order to implement proper frame-alignment. We also have the following equations, which can instead be evaluated for arbitrary $T$, subject to $|\Omega| T\gg 1$:
\begin{equation}
    \ln\left[\frac{2}{\widehat{\braket{\sigma_z(T)}}^{\Omega_z}_{z_+}-\widehat{\braket{\sigma_z(T)}}^{\Omega_z}_{z_-}}\right]\!=S_{1,-1}^+(\Omega+\omega_q)T -\frac{1}{2}\ln[\alpha],
\label{eq::multispamfirst}
\end{equation}
\begin{equation}
\label{eq::SPAMczdrm}
    \ln\left[\frac{2}{\widehat{\braket{\sigma_z(T)}}^{\!-\Omega_z}_{z_+}-\widehat{\braket{\sigma_z(T)}}^{\!-\Omega_z}_{z_-}}\right]\!=\!S_{-1,1}^+(\Omega-\omega_q)T    -\frac{1}{2}\ln[\alpha],
\end{equation}
\begin{equation}
    \ln\left[\frac{2}{\widehat{\braket{\sigma_x(T)}}^{\Omega_x}_{x_+}-\widehat{\braket{\sigma_x(T)}}^{\Omega_x}_{x_-}}\right]\!=\!A(\Omega)T-\frac{1}{2}\ln\left[\alpha\right].\, \label{eq::classdephspam}
\end{equation}
These equations show that linear regression on experimental estimates on the left hand-side of Eqs.\,\eqref{eq::dephspamfirst}-\eqref{eq::classdephspam} yields a SPAM-free estimate of the target classical spectra as the slope and the SPAM parameters as the intercepts.

\noindent
\begin{table*}[t]
\setlength{\fboxrule}{0.8pt}
\fbox{
\begin{minipage}[t][17.2cm]{0.8\textwidth} 
\raggedright 
        \mbox{\bf{Protocol 4: SPAM-robust multi-axis spin-locking QNS} }\\
        \rule{14.2cm}{0.5pt}
        \noindent \mbox{\textit{Input.} Driving amplitude $\Omega$, a set of times $\{T_j\}$.}\\
        \vspace*{1mm}
        \noindent \mbox{\textit{Output.} 
        Noise spectra at frequency $\Omega$ and SPAM parameters.} \\     
        \vspace*{1mm}   
        \noindent \mbox{\textbf{The protocol:} \hspace*{2cm}     }  
\begin{enumerate}
\item[{\bf (1)}] Implement single-axis QNS with $H_\text{ctrl} = \tfrac{1}{2}\Omega\sigma_z$:
\vspace*{-0mm}
\begin{enumerate}
\item[\textbf{a)}] Prepare the initial state $\ket{z_+}\bra{z_+}$, apply a resonant drive along $z$ with amplitude $\Omega$. Let the qubit evolve for a set of times $\{T_j\}$, with $|\Omega| T_j\gg 1$, and measure $\sigma_z$.
\item[\textbf{b)}] Repeat \textbf{a)} sufficiently many times to estimate $\{\widehat{\braket{\sigma_z(T_j)}}^{\Omega_z}_{z_+}\}$.
\item[\textbf{c)}] Repeat steps \textbf{a)} and \textbf{b)} with the qubit prepared in $\ket{z_-}\bra{z_-}$, to estimate $\{\widehat{\braket{\sigma_z(T_j)}}_{z_-}^{\Omega_z}\}$.
\item[\textbf{d)}] Prepare the initial state $\ket{x_+}\bra{x_+}$, apply a resonant drive along $z$ with amplitude $\Omega$. Let the qubit evolve for a set of times $\{T^{(n)}_j\}$, with $|\Omega| T^{(n)}_j\gg 1$, $T^{(n)}_j = 2\pi n_j/|\Omega|$, $n_j\in {\mathbb N}$, and measure $\sigma_z$. 
\item[\textbf{e)}] Repeat \textbf{d)} sufficiently many times to estimate $\{ \widehat{ \langle \sigma_x(T^{(n)}_j) \rangle}_{x_+}$ $\hspace*{-4mm}^{\Omega_z}\}$.

\item[\textbf{f)}] Repeat steps \textbf{d)} and \textbf{e)} with the qubit prepared in $\ket{x_-}\bra{x_-}$, to estimate 
$\{ \widehat{ \langle \sigma_x(T^{(n)}_j) \rangle}_{x_-}$ $\hspace*{-4mm}^{\Omega_z}\}$.
\end{enumerate}

\item[{\bf (2)}] Use the expectation values and Eqs.\,\eqref{eq::multispamfirst}, \eqref{eq::dephspamfirst}, \eqref{eq::SPAMqzdrp} to estimate $S_{1,-1}^+(\Omega+\omega_q)$, $S_{-1,1}^-(-\Omega-\omega_q)$ and $S_{0,0}(0)$, along with the SPAM parameters $\alpha_M$, $\delta$ by approximating $\alpha\approx\alpha_M$.

\item[{\bf (3)}] Implement single-axis QNS with $H_\text{ctrl} = -\tfrac{1}{2}\Omega\sigma_x$:
\begin{enumerate}
\item[\textbf{a)}] Prepare the initial state $\ket{z_+}\bra{z_+}$, apply a resonant drive along $z$ with amplitude $\Omega$. Let the qubit evolve for 
 a set of times $\{T_j\}$ such that $|\Omega| T_j \gg 1$, and measure $\sigma_z$.
\item[\textbf{b)}] Repeat \textbf{a)} sufficiently many times to estimate $\{\widehat{\braket{\sigma_z(T_j)}}^{-\Omega_z}_{z_+}\}$.
\item[\textbf{c)}] Repeat steps \textbf{a)} and \textbf{b)} with the qubit prepared in $\ket{z_-}\bra{z_-}$ to estimate $\{\widehat{\braket{\sigma_z(T_j)}}_{z_-}^{-\Omega_z}\}$.
\end{enumerate}
\item[{\bf (4)}] Use the expectation values and Eqs.\,\eqref{eq::SPAMczdrm}, \eqref{eq::SPAMqzdrm} to estimate $S_{- 1 , 1}^+(\Omega-\omega_q)$ and $S_{ 1, - 1}^-(-\Omega+\omega_q)$.

\item[{\bf (5)}] Implement single-axis QNS with $H_\text{ctrl} = \tfrac{1}{2}\Omega\sigma_x$:
\vspace*{-0mm}
\begin{enumerate}
\item[\textbf{a)}] Prepare the initial state $\ket{x_+}\bra{x_+}$, apply a resonant drive along $x$ with amplitude $\Omega$. Let the qubit evolve for a set of times $\{T_j\}$ such that $|\Omega| T_j\gg 1$, and measure $\sigma_x$.
\item[\textbf{b)}] Repeat \textbf{a)} sufficiently many times to estimate $\braket{\sigma_x(T_j)}^{\Omega_x}_{x_+}$.
\item[\textbf{c)}] Repeat steps \textbf{a)} and \textbf{b)} with the qubit prepared in $\ket{x_-}\bra{x_-}$, to estimate $\braket{\sigma_x(T_j)}_{x_-}^{\Omega_x}$.
\end{enumerate}
\item[\bf{(6)}] Fit the data to Eqs.\,\eqref{eq::classdephspam}-\eqref{eq::SPAMqxdr} to estimate $A(\Omega)$, $B(\Omega)$, and use the estimated transverse spectrum to infer the dephasing classical and quantum spectrum:
\begin{itemize}
    \item $S^+_{0,0}(\Omega)= A(\Omega)-\frac{1}{2}\left[S^+_{1,-1}(\Omega+\omega_q)+S^+_{-1,1}(\Omega-\omega_q)\right]$,
    \item $S^-_{0,0}(\Omega)= B(\Omega)-\frac{1}{2}\left[S^-_{1,-1}(-\Omega+\omega_q)-S^-_{-1,1}(-\Omega-\omega_q)\right]$.
\end{itemize}    
\end{enumerate}
\end{minipage} 
}
\caption{Summary of the proposed SPAM-robust SL QNS protocol for multi-axis noise.} 
\label{Protocol4}
\end{table*}

\subsection{Determining SPAM-free quantum spectra}

The experimentally estimated quantities that relate to the quantum spectrum can be obtained via a procedure similar to the one leading to Eq.\,\eqref{eqn::quantnonlinear} for single-axis noise, repeated, however, for each of the three relevant drives. We obtain:
\begin{multline}
    \label{eq::SPAMqzdrp}\frac{\widehat{\braket{\sigma_z(T)}}_{z_+}^{\Omega_z}+\widehat{\braket{\sigma_z(T)}}_{z_-}^{\Omega_z}}{2}=\\
    \alpha_M \frac{S_{-1, 1}^-(-\Omega-\omega_q)}{S_{1, -1}^+(\Omega+\omega_q)}\left(e^{-2S_{1, -1}^+(\Omega+\omega_q)T}- 1\right) + \delta,
\end{multline}
\begin{multline}
    \label{eq::SPAMqzdrm}\frac{\widehat{\braket{\sigma_z(T)}}_{z_+}^{\!-\Omega_z}+\widehat{\braket{\sigma_z(T)}}_{z_-}^{\!-\Omega_z}}{2}=\\
    \alpha_M \frac{S_{ 1, - 1}^-(-\Omega+\omega_q)}{S_{- 1 , 1}^+(\Omega-\omega_q)}\left(1 - e^{-2S_{- 1,  1}^+(\Omega-\omega_q)T}\right) + \delta,
\end{multline}
\begin{align}
   \label{eq::SPAMqxdr} \frac{\widehat{\braket{\sigma_x(T)}}_{x_+}^{\Omega_x}+\widehat{\braket{\sigma_x(T)}}_{x_-}^{\Omega_x}}{2}&=\alpha_M \frac{B(\Omega)}{A(\Omega)} \left(1-e^{-A(\Omega)T} \right)+\delta.
\end{align}
Altogether, Eqs.\,(\ref{eq::multispamfirst})-(\ref{eq::SPAMqxdr}) provide a system of equations that allows us to estimate \emph{all} the SPAM-free classical spectra, and the quantum spectra up to a scaling factor in $\alpha_M$, namely, $\alpha_M S^-_{1,-1}(-\Omega +\omega_q)$, $\alpha_M S^-_{-1,1}(-\Omega -\omega_q)$ and $\alpha_M S^-_{0,0}(\Omega)$. Similarly to the single-axis noise scenario, we need to either estimate the measurement error parameter $\alpha_M$ independently of $\alpha_{SP}$, or work in a regime where the measurement errors are the dominant source of SPAM errors ($\alpha\approx\alpha_M$). Here, we make the same assumption made for the single-axis SPAM-robust SL QNS, that $\alpha\approx\alpha_M$.

Our complete protocol for multi-axis SPAM-robust QNS may be summarized as in Table\,\ref{Protocol4}.

\subsection{Protocol validation: IBM Q results}
\label{sec::multiqnsprotocol}

\begin{figure*}
    \centering
    \includegraphics[scale=0.8]{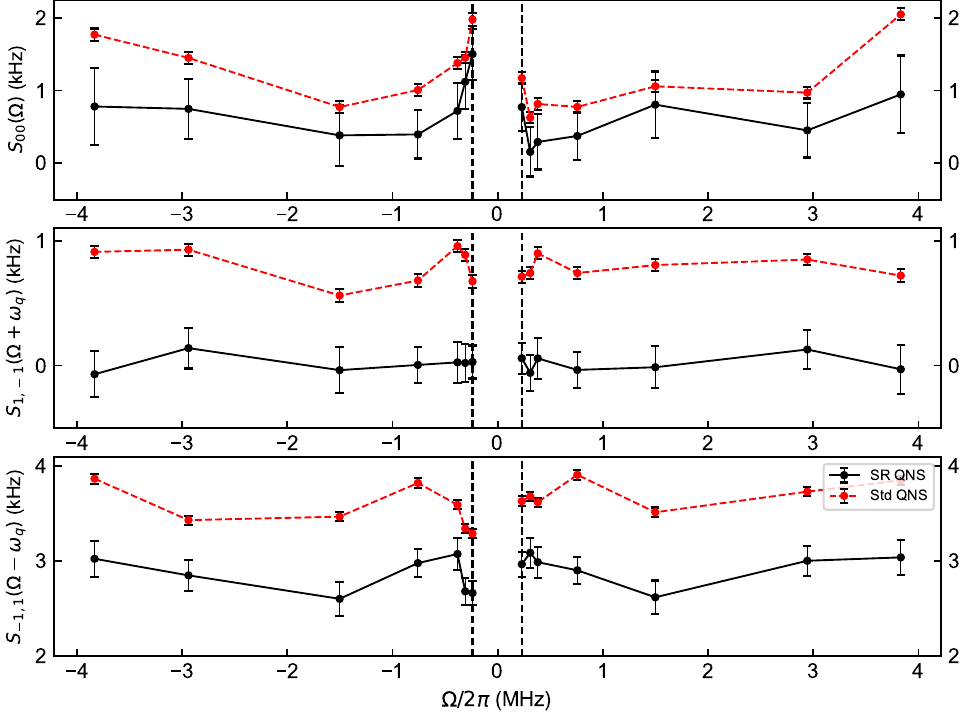}
    \vspace*{-2mm}
    \caption{(Color online) {\bf SPAM-robust QNS IBM Q reconstructions.} Complete spectral reconstructions using standard multi-axis SL QNS protocol (red) and the SPAM-robust QNS protocol (black). Each experimental run was taken over $2,000$ shots, and the error bars represent a $95\%$ confidence interval. }
    \label{fig:4}
\end{figure*}

In an effort to validate the SPAM-robust SL QNS protocol we just discussed, we conducted cloud-based simulations to implement it on IBM Q, by using a single-qubit device ({\em ibmq\_armonk}, now-retired). While this qubit reported a higher percentage of measurement errors and measurement error asymmetries than current devices do, the fact that no additional qubits were present made it a convenient testbed for benchmarking our protocol in the absence of complicating factors stemming from qubit crosstalk. 

For this qubit, the reported frequency $\nu_q\equiv \omega_q / (2\pi)= 4.97\,\mathrm{GHz}$, with nominal relaxation and dephasing times $T_1=169.66\,\mu s$ and $T_2=256.44\,\mu s$, respectively. Two main modifications were needed to accommodate practical constraints. First, continuous drives along the $z$-axis were implemented using ``virtual $z$-gates'' followed by idling the qubit for a short time. Second, continuous drives along the $x$-axis were implemented as a train of 
pulses of constant amplitude, to circumvent some limitations to the control input at the Application Programming Interface (API; see Appendix\,\ref{app::constdrive} for details). 

In addition to the above, due to limited available run-time on the cloud-based platform, it was not possible to execute the complete multi-axis SPAM-robust QNS protocol described in Table\,\ref{Protocol4}. The observable data we collected consisted of 
\begin{eqnarray*}
\{\widehat{\braket{\sigma_z(T_{j})}}^{\Omega_z}_{z_\pm}\}, \;
\{\widehat{\braket{\sigma_z(T_{j})}}^{-\Omega_z}_{z_\pm}\}, \;
\{\widehat{\braket{\sigma_x(T_{j})}}^{\Omega_x}_{x_\pm}\}, 
\end{eqnarray*}
for a set of $14$ drive amplitudes. However, we omitted the measurement of 
$\{ \widehat{ \langle \sigma_x(T^{(n)}_j) \rangle}_{x_\pm}$ $\hspace*{-4mm}^{\Omega_z}\}$, meaning that we could not characterize $S_{0,0}(0)$. In all experiments, we chose the driving field amplitudes in the range $-3.83\, \mathrm{MHz}\lesssim\Omega\lesssim 3.83\,\mathrm{MHz}$. Note that, in all the reconstructions, we excluded a low-frequency window, $|\Omega| \lesssim 2.3\,\mathrm{kHz}$, due to the fact that the data becomes dominated by oscillations, as also reported and discussed in previous SL experiments \cite{Yan2013}. While our limited access to the system prevented a conclusive assessment, the most likely candidate for these oscillations appears to be dephasing taking place between between the ramp-up and ramp-down of pulses, which could be in principle mitigated by more elaborated SL sequences \cite{Yan2013}.

\subsubsection{Reconstructed noise spectra}

Figure\,\ref{fig:4} summarizes representative results on our spectral reconstructions. Correcting for SPAM errors paints quite a different picture of the noisy environment for this device. Altogether, it shows that the actual noise that the qubit experiences is \emph{measurably smaller} than we may infer using standard QNS techniques. In particular, our data shows that the spherical excitation spectrum, $S_{1,-1}(\Omega+\omega_q)$, vanishes after the effects of SPAM-errors are removed. Thus, it appears as if SPAM is contributing to spurious excitation of the qubit. This confirms how, physically, SPAM-errors manifest as bias to the true noise spectrum, consistent with our theoretical and numerical simulations (for dephasing noise). 

Fig.\,\ref{fig:4} also shows the dephasing spectrum $S_{0,0}(\Omega)$, inferred through the non-robust multi-axis SL QNS protocol, has an increasing trend at higher drive amplitudes, in contrast with the nearly-plateau level it reaches when estimated through the SPAM-robust protocol. This behavior may be seen as an artifact of the standard protocols having been implemented using shorter evolution times $T$ for higher driving frequencies (while maintaining the SL condition), due to practical limitations. As seen from Eqs.\,\eqref{eq::wings1}-\eqref{eq::wings2}, for shorter evolution times, the contribution of SPAM errors becomes more prominent. By design, this effect has {\em no} bearing on the SPAM-robust QNS protocol and is thus not visible there. Also notable is the fact that, while the transverse spectra appear to be nearly white apart from some features around lower frequencies, the dephasing spectrum has visible \emph{asymmetries} in the positive vs. negative frequency axis. This indicates the presence of \emph{non-classical} noise contributing to the spectrum in this domain. 

\begin{figure}[t!]
    \centering
    \includegraphics[width=0.45\textwidth]{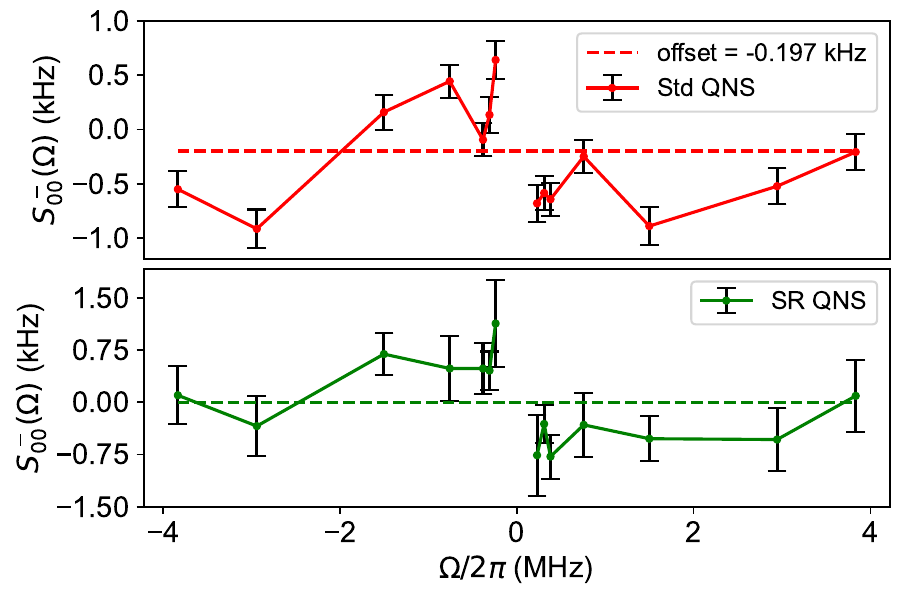}
    \vspace*{-2mm}
    \caption{(Color online) {\bf IBM Q dephasing quantum spectrum.} 
    The top plot shows (in red) the dephasing spectrum reconstructed from the standard QNS, whereas the bottom plot shows (in green) the SPAM-robust reconstruction. These plots are constructed from the same data as Fig.\,\ref{fig:4}. The offset in the top plot can be directly attributed to the measurement asymmetry parameter $\delta$. The SPAM-robust QNS removes this offset, making the quantum spectrum anti-symmetric, as expected on general grounds.}
    \vspace*{-4mm}
    \label{fig:5}
\end{figure}

This conclusion is further reinforced by examining the quantum dephasing spectrum $S^-_{0,0}(\Omega)$, which is displayed in  
Fig.\,\ref{fig:5}. In the top-plot, the spectrum is not anti-symmetric as inferred through the uncorrected multi-axis SL QNS protocol. Rather, it demonstrates an offset from the $S^-_{0,0}(\Omega)=0$ line that we expect from Eq.\,\eqref{eq::SPAMqxdr}, corresponding to the the asymmetry parameter $\delta$. This offset is not present in the bottom plot, where the spectrum is inferred using our modified SPAM-robust SL QNS protocol.

\begin{figure*}
    \centering
    \includegraphics[scale=0.8]{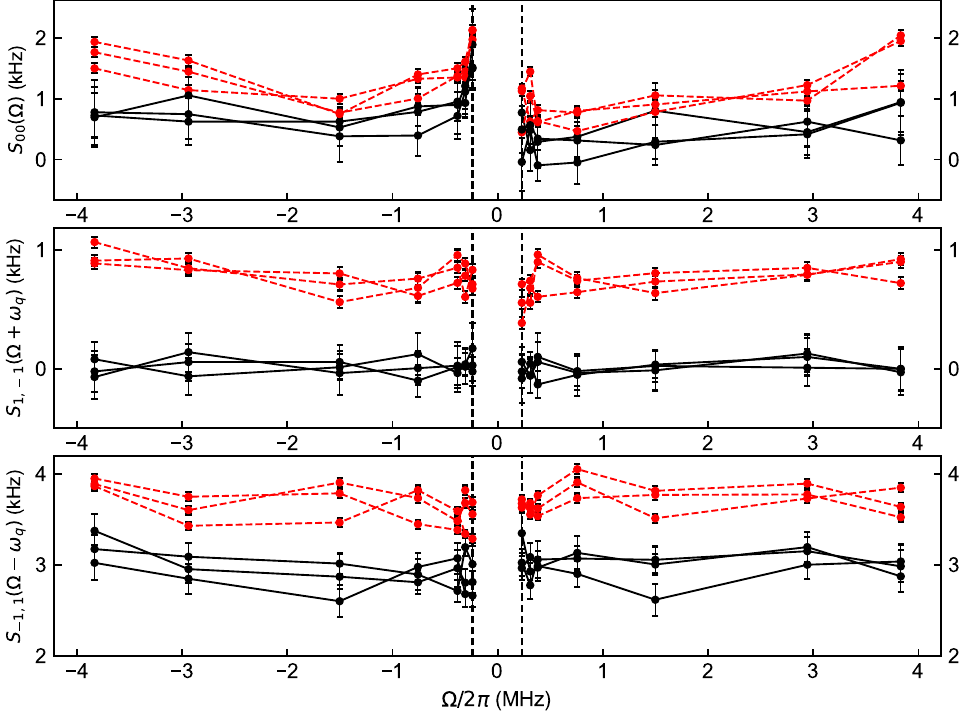}
    \vspace*{-2mm}
    \caption{(Color online) {\bf Comparison between IBM Q QNS reconstructions.} Spectral reconstruction results for three experiments carried out within the same calibration cycle over a period of 14-hours. The standard QNS (red) and the SPAM-robust QNS (black) both show consistency and the qualitative features of the spectra do not change. The top-panel is the dephasing noise spectrum, the middle-panel is the excitation transverse spectrum and the bottom-panel is the relaxation transverse spectrum.}
    \label{fig:6}
\end{figure*}

An important question that arises in the context of QNS is the extent to which the noise processes may be taken to be stationary, and thus characterization results consistent over a significant period of time. To address this question, we ran the multi-axis SPAM-robust SL QNS protocol three times over a period of 14 hours within one calibration cycle (of 24 hours). As Fig.\,\ref{fig:6} shows, both the standard and the SPAM-robust spectral reconstructions are remarkable consistency over a few hours, suggesting a high degree of stationarity for noise sources in this qubit device.

\subsubsection{SPAM parameters}

In addition to the noise spectra, under our working assumption of negligible SP errors, the SPAM-robust QNS protocol also allows us to estimate the SPAM parameters $\alpha_M$ and $\delta$. Table\,\ref{Table:1} shows the characterized SPAM-parameters and compares them to the values reported by IBM. As one can see, the reported value of $\alpha_M$ from the calibration data agrees remarkably well and falls within the error bars of the estimate obtained from the SPAM-robust SL QNS protocol. However, the value of the parameter $\delta$ in the calibration data falls outside the $95\%$ confidence-interval of the estimated value. While no error bars are available for the calibration data, we expect that taking data for a large set of evolution times could improve the estimate.

\begin{table}
\caption{Estimated values for the SPAM parameters at a $95\%$ confidence interval. All values should be multiplied by $10^{-2}$.}
\vspace*{0.5mm}
\begin{tabular}{|c|r|r|r|c|}
 \hline
 \multicolumn{5}{|c|}{{\bf SPAM Parameters}} \\
 \hline\hline
 \label{Table:1}Parameter & Expt 1$\quad$
 & Expt 2$\quad$ 
 & Expt 3$\quad$ 
 & Calibration data \\
  \hline
 $\alpha_M$ &$87.2\pm1.1$&$88.3\pm1.1$& $87.2\pm1.2$& 88.0\\
 $\delta$ &$3.8\pm1.4$&$3.2\pm1.4$&$3.6\pm1.5$& 1.38\\
 \hline
\end{tabular}
\end{table}

We also note that the estimated parameters from the SPAM-robust QNS protocol can be directly related to the measurement error mitigation routines based on ``confusion matrices'' that are commonly employed \cite{PhysRevA.103.042605}, especially by IBM Q, to calibrate the systems. While a challenge of this approach is that the matrix $A$'s invertibility cannot be guaranteed, our protocol provides an alternative procedure, where SPAM parameters are inferred ``self-consistently'', along with spectral information, and invertibility is ensured by construction.

\section{Discussion and outlook}
\label{sec::discussion}

We have developed the theoretical framework for a quantum noise spectroscopy protocol able to characterize competing transverse and dephasing temporally correlated noise processes on a qubit, and further shown how the resulting protocol can be modified so it is immune to a large class of static SPAM errors. Our protocol relies upon, and further extends, the use of continuous control characterization methods inspired to spin-locking relaxometry. We have demonstrated our protocol in action by reconstructing the multi-axis noise spectra of a single transmon qubit on the IBM Quantum Platform. Our results show how, even in the simplest setting of a single-qubit, SPAM errors can lead to a {\em significant mischaracterization} of the target noise spectra -- by causing, in particular, both dephasing and transverse spectra to be up-shifted in value and quantum spectra to lack the requisite spectral asymmetry properties.

Our proposed SPAM-robust single- and multi-axis QNS protocols provide a valuable and timely addition to the existing toolbox for noise characterization. With appropriate modifications, we expect that they may be further improved to also incorporate robustness against leading control errors, via the use of enhanced spin-locking sequences \cite{Yan2013}, and possibly extended to multi-level systems \cite{Sung2021-rk}. 

Yet, important limitations remain to be addressed. As presented, our protocols fall short of characterizing the state-preparation and measurement errors separately and rely on one of them being the dominant source of SPAM-errors. This limitation is intrinsic to the chosen sensing modality, as we have shown that it is not possible to determine all the SPAM-parameters through continuously-driven single-qubit dynamics. It remains open to determine whether more powerful QNS schemes may be constructed, that could afford to disambiguate between these two SPAM sources -- possibly by augmenting the available time-dependent control and/or measurement resources. It is important to note that this limitation does not challenge the SPAM robustness of the protocol. If state-preparation errors for the sensor can be characterized independently \cite{Mattias, Laflamme2021, Wei2023}, the SPAM robust QNS can accurately estimate the complete set of spectra for competing state-preparation and measurement errors. 

Another limitation stems from the fact that our approach relies on truncating a time-convolutionless master equation to the second order. As such, it is not suitable for characterizing noise with a non-Gaussian statistics, for which higher-order spectral estimation is needed. Although weak in comparison to the Gaussian contribution, non-Gaussian noise effects may lead to unique dynamical signatures that cannot captured by spectra alone \cite{PhysRevLett.116.150503,Sung2019}. SPAM-robust methods could thus play an instrumental role in enabling the leading higher-order spectra to be accurately characterized. A promising venue in this respect could be to explore whether SPAM-robust modifications of frame-based QNS protocols may be devised \cite{PazSilvaFrames}, and then extended to the non-Gaussian regime by building on recent work \cite{Wenzheng}.

Finally, extending the reach of SPAM-robust QNS beyond a single qubit remains an outstanding challenge. Two-qubit systems in a dephasing-dominated regime could provide a tractable yet relevant setting for investigating SPAM-robust extensions of existing spin-locking \cite{Uwe2020} or, possibly, pulsed comb-based QNS approaches \cite{PhysRevA.95.022121}. Notably, accurate characterization of spatio-temporal noise correlations could prove vital for further boosting the fidelity of entangling two-qubit gates in spin qubits \cite{Dzurak} through noise-tailored optimal design. We leave this to a future, separate investigation.  

\section*{Acknowledgements}

It is a pleasure to thank Gerardo A. Paz-Silva and F\'elix B\'eaudoin for enlightening discussions and early contributions to the problem of multi-axis SL QNS. We also thank Vivian Maloney, Aaron Kleger, and Mattias Fitzpatrick for valuable input on various aspects of the project. This work was supported by the U.S. Department of Energy, Office of Science, Office of Advanced Scientific Computing Research, Accelerated Research in Quantum Computing under Award No.\,DE-SC0020316 and AW-D00011587. Work at Dartmouth was also supported in part by the U.S. Army Research Office through Grant No.\,W911NF-18-1-0218 and Grant No.\,W911NF-22-1-0004. This research used resources of the Oak Ridge Leadership Computing Facility, which is a DOE Office of Science User Facility supported under Contract DE-AC05-00OR22725. We acknowledge the use of IBM Quantum services for this work. The views expressed are those of the authors, and do not reflect the official policy or position of IBM or the IBM Quantum team.


\appendix

\begin{widetext}
\section{TCL master equation for non-Markovian multi-axis quantum noise}
\label{appendix::tclmema}

To determine the controlled qubit dynamics under multi-axis quantum noise in the toggling-frame of the control, the first step is to substitute the spherical-basis toggling frame Hamiltonian in Eq.\,\eqref{eq::HTogglingMulti} into the second-order TCL ME in Eq.\,\eqref{eq::TCL2ndOrder}. As in the main text, we drop the tilde notation, which leads to 
\begin{align*}
    \dot{\rho}(t)=
    -\sum_{\substack{\alpha,\beta,\alpha',\beta'\\=-1,0,1}}\,\int_0^t\!\!ds\;e^{i\alpha\omega_q t}e^{i\alpha'\omega_q s}\,y_{\alpha\beta}(t)y_{\alpha'\beta'}(s)\,\Big\langle\,\Big[\sigma_\beta\otimes B_{-\alpha}(t),\Big[\sigma_{\beta'}\otimes B_{-\alpha'}(s),\rho(t)\otimes\rho_B\Big]\Big]\,\Big\rangle.
\end{align*}
We can then determine the evolution of the density matrix elements in an arbitrary qubit basis $\{ \ket{u_j} \}$, with $u=x,y,z$, and $j=\pm$ (Eq.\,\eqref{eigs}), by evaluating
\begin{align*}
\dot{\rho}_{u_j u_{j'}} 
= &-\sum_{\ell=\pm1}\;\sum_{\substack{\alpha,\beta,\alpha',\beta'\\=-1,0,1}}\;\int_0^t\!\!ds\,e^{i\alpha \omega_q t}e^{i\alpha' \omega_q s}y_{\alpha\beta}(t)y_{\alpha'\beta'}(s)\,
\Big[\expect{B_{-\alpha'}(s)B_{-\alpha}(t)}\,\bra{u_\ell}\sigma_{\beta'}\sigma_{\beta}\ket{u_{j'}}\rho_{u_j u_\ell}\\&\quad\quad\quad\quad\quad\quad\quad\quad\quad\quad\quad\quad
\quad\quad\quad\quad\quad\quad\quad\quad\quad
+\expect{B_{-\alpha}(t)B_{-\alpha'}(s)}\,\bra{u_j}\sigma_{\beta}\sigma_{\beta'}\ket{u_\ell}\rho_{u_\ell u_{j'}}\Big]\\
&+\sum_{\ell,\ell'=\pm 1}\;\sum_{\substack{\alpha,\beta,\alpha',\beta'\\=-1,0,1}}\;\int_0^t\!\!ds\,e^{i\alpha \omega_q t}e^{i\alpha' \omega_q s}y_{\alpha\beta}(t)y_{\alpha'\beta'}(s)
\Big[\expect{B_{-\alpha'}(s)B_{-\alpha}(t)}\,\bra{u_\ell}\sigma_{\beta'}\ket{u_{j'}}\bra{u_j}\sigma_{\beta}\ket{u_{\ell'}}\\&\quad\quad\quad\quad\quad\quad\quad\quad\quad\quad\quad
\quad\quad\quad\quad\quad\quad\quad\quad\quad\quad
+\expect{B_{-\alpha}(t)B_{-\alpha'}(s)}\,\bra{u_\ell}\sigma_{\beta}\ket{u_{j'}}\bra{u_j}\sigma_{\beta'}\ket{u_{\ell'}}\Big]\rho_{u_{\ell'}u_{\ell}},
\end{align*}
where we have inserted resolutions of the identity $\mathcal{I}=\sum_{\ell=\pm 1}|u_\ell\rangle\langle u_\ell |$ and, to reduce clutter, we have suppressed the time argument from the density operator, $\rho_{u_ju_{j'}}(t)=\rho_{u_ju_{j'}}$.

To make contact with the spherical spectra, we transform this expression to the frequency domain,
\begin{align}
\label{eq::drho1}
&\dot\rho_{u_ju_{j'}}=-\frac{1}{2\pi}\sum_{\ell=\pm1}\;\sum_{\substack{\alpha,\beta,\alpha',\beta'\\=-1,0,1}}e^{i(\alpha+\alpha')\omega_qt}\!\int_{-\infty}^\infty\!\!\!d\omega\,e^{i\omega t}y_{\alpha\beta}(t)\!\!\int_0^t\!\!ds\,e^{-i\omega s}y_{\alpha'\beta'}(s)\\&\quad\quad\quad\quad\quad\quad\quad
\Big[S_{-\alpha',-\alpha}(-\omega-\alpha'\omega_q)\bra{u_\ell}\sigma_{\beta'}\sigma_{\beta}\ket{u_{j'}}\rho_{u_j u_\ell}
+S_{-\alpha,-\alpha'}(\omega+\alpha'\omega_q)\bra{u_j}\sigma_{\beta}\sigma_{\beta'}\ket{u_\ell}\rho_{u_\ell u_{j'}}\Big]\notag\\\notag
&+\frac{1}{2\pi}\sum_{\ell,\ell'=\pm1}\;\sum_{\substack{\alpha,\beta,\alpha',\beta'\\=-1,0,1}}e^{i(\alpha+\alpha')\omega_qt}\!\!\int_{-\infty}^\infty\!\!\!d\omega\,e^{i\omega t}y_{\alpha\beta}(t)\!\!\int_0^t\!\!ds\,e^{-i\omega s}y_{\alpha'\beta'}(s)
\\&\quad\quad\quad
\Big[S_{-\alpha',-\alpha}(-\omega-\alpha'\omega_q)\bra{u_\ell}\sigma_{\beta'}\ket{u_{j'}}\bra{u_j}\sigma_{\beta}\ket{u_{\ell'}}
+S_{-\alpha,-\alpha'}(\omega+\alpha'\omega_q)\bra{u_\ell}\sigma_{\beta}\ket{u_{j'}}\bra{u_j}\sigma_{\beta'}\ket{u_{\ell'}}\Big]\rho_{u_{\ell'}u_\ell}.\notag
\end{align}
Note that this expression contains an exponential $e^{i(\alpha+\alpha')\omega_qt}$, which is rapidly oscillating when $\omega_qt\gg1$ and $\alpha+\alpha'\neq 0$. To obtain Eq.\,\eqref{eq::SecularApprox} in the main text,
 we made the secular approximation and discarded all rapidly oscillating terms with $\alpha+\alpha'\neq 0$.

\subsection{Derivation via multi-scale perturbation theory}
\label{app:multiscale}

The secular approximation, which holds in the limit $\omega_qt\gg1$, can be formally derived as the leading-order term in a multi-scale perturbative expansion of the TCL ME \cite{Kevorkian1996,Diego}. While the estimation procedure in the main text relies on the secular approximation, we include the multi-scale expansion here for the purpose of obtaining higher-order dynamical corrections. Incorporating these higher order corrections, in principle, allows the ME to be extended beyond the $\omega_qt\gg1$ regime.

Let $\omega_q\equiv \omega_0\gamma$, where $\omega_0$ has units of frequency and $\gamma\gg 1$ is dimensionless. By examining Eq.\,\eqref{eq::drho1}, we see that it depends on two timescales, a ``slow" time $t$ and a ``fast" time, $\tau\equiv\gamma t$. Accordingly, we can write \erf{eq::drho1} using the abbreviated notation
\begin{align*}
\dot{\rho}_{u_ju_{j'}}=\!\!\!
\sum_{\alpha,\alpha'= \pm 1 } \!\!e^{i(\alpha+\alpha')\omega_0\tau}
\!\bigg\{\sum_{\ell=\pm 1}\Big[A_{u_ju_{j'}}(\alpha,\alpha',\ell,t)\rho_{u_ju_\ell}+B_{u_ju_{j'}}(\alpha,\alpha',\ell,t)\rho_{u_\ell u_{j'}}\Big]
+\!\!\sum_{\ell,\ell'=\pm 1}C_{u_ju_{j'}}(\alpha,\alpha',\ell,\ell',t)\rho_{u_{\ell'}u_\ell}\bigg\}.
\end{align*}
Note that $A_{u_ju_{j'}}(\alpha,\alpha',\ell,t)$, $B_{u_ju_{j'}}(\alpha,\alpha',\ell,t)$, and $C_{u_ju_{j'}}(\alpha,\alpha',\ell,\ell',t)$ depend only on the slow time.

The method of multiple scales involves converting this system of ordinary DEs in the slow time $t$ into a system of partial DEs in $t$ and $\tau$. Using the chain rule, we can write the total derivative with respect to $t$ above as
\begin{align}\notag
\dot\rho_{u_ju_{j'}}=\,\frac{\partial\rho_{u_ju_{j'}}}{\partial t}+\gamma\frac{\partial\rho_{u_ju_{j'}}}{\partial \tau}
=&\sum_{\alpha,\alpha'=\pm 1}e^{i(\alpha+\alpha')\omega_0\tau}\bigg\{\sum_{\ell=\pm 1}\Big[A_{u_ju_{j'}}(\alpha,\alpha',\ell,t)\rho_{u_ju_\ell}+B_{u_ju_{j'}}(\alpha,\alpha',\ell,t)\rho_{u_\ell u_{j'}}\Big] \notag \\
&\quad\quad\quad\quad\quad\quad\quad\;\;+\sum_{\ell,\ell'=\pm 1}C_{u_ju_{j'}}(\alpha,\alpha',\ell,\ell',t)\rho_{u_{\ell'}u_{\ell}}\bigg\},
\label{eq::drho3}
\end{align}
where each density matrix element depends implicitly on both the slow and the fast times. We seek a set of solutions for the density matrix elements satisfying the Ansatz 
\begin{align}
\label{eq::ansatz}
\rho_{u_ju_{j'}}(t,\tau)=\rho_{u_ju_{j'}}(t)+\frac{1}{\gamma}\,\xi_{u_ju_{j'}}(t,\tau),
\end{align}
with initial conditions $\rho_{u_ju_{j'}}(0)=\bra{u_j}\rho(0)\ket{u_{j'}}$ and $\xi_{u_ju_{j'}}(0,\tau)=\xi_{u_ju_{j'}}(t,0)=0$. 
By substituting the Ansatz for the density matrix elements in Eq.\,\eqref{eq::drho3}, we obtain
\begin{align*}
\frac{\partial \rho_{u_ju_{j'}}}{\partial t}\,+\,&\frac{1}{\gamma}\,\frac{\partial\xi_{u_ju_{j'}}(t,\tau)}{\partial t}+\frac{\partial\xi_{u_ju_{j'}}(t,\tau)}{\partial \tau}\\
=&\sum_{\alpha,\alpha'=\pm 1}e^{i(\alpha+\alpha')\omega_0\tau}\bigg\{\sum_{\ell=\pm 1}\Big[A_{u_ju_{j'}}(\alpha,\alpha',\ell,t)\rho_{u_ju_\ell}(t)+B_{u_ju_{j'}}(\alpha,\alpha',\ell,t)\rho_{u_\ell u_{j'}}\Big]\\
&\quad\quad\quad\quad\quad\quad\quad\quad\quad\;\;+\sum_{\ell,\ell'=\pm 1}C_{u_ju_{j'}}(\alpha,\alpha',\ell,\ell',t)\rho_{u_{\ell'}u_\ell}\bigg\}\\
&+\frac{1}{\gamma}\sum_{\alpha,\alpha'=\pm 1}e^{i(\alpha+\alpha')\omega_0\tau}\bigg\{\sum_{\ell=\pm 1}\Big[A_{u_ju_{j'}}(\alpha,\alpha',\ell,t)\xi_{u_ju_\ell}(t,\tau)+B_{u_ju_{j'}}(\alpha,\alpha',\ell,t)\xi_{u_\ell u_{j'}}(t,\tau)\Big] \\
&\quad\quad\quad\quad\quad\quad\quad\quad\quad\;\;
+\sum_{\ell,\ell'=\pm 1}C_{u_ju_{j'}}(\alpha,\alpha',\ell,\ell',t)\xi_{u_{\ell'}u_{\ell}}(t,\tau)\bigg\}.
\end{align*}
To arrive at a perturbative expansion for each partial derivative in the system, we replace $\partial \rho_{u_ju_{j'}}(t)/\partial t$ and $\xi_{u_ju_{j'}}(p,t,\tau)$ with asymptotic expansions in $(1/\gamma)$,
\begin{align*}
&\frac{\partial \rho_{u_ju_{j'}}}{\partial t}=\eta_{u_ju_{j'}}^{(0)}(t)+\frac{1}{\gamma}\,\eta_{u_ju_{j'}}^{(1)}(t)
+\frac{1}{\gamma^2}\,\eta_{u_ju_{j'}}^{(2)}(t)+\ldots\\
&\xi_{u_ju_{j'}}(t,\tau)=\xi_{u_ju_{j'}}^{(0)}(t,\tau)+\frac{1}{\gamma}\,\xi_{u_ju_{j'}}^{(1)}(t,\tau)
+\frac{1}{\gamma^2}\,\xi_{u_ju_{j'}}^{(2)}(t,\tau)+\ldots\;\;.
\end{align*}
After rearranging the terms, we obtain
\begin{align*}
0=&\big(\eta_{u_ju_{j'}}^{(0)}(t)+\frac{1}{\gamma}\eta_{u_ju_{j'}}^{(1)}(t)+\ldots\big)+\frac{1}{\gamma}\frac{\partial}{\partial t}
\big(\xi_{u_ju_{j'}}^{(0)}(t,\tau)+\frac{1}{\gamma}\xi_{u_ju_{j'}}^{(1)}(t,\tau)+\ldots\big)
+\frac{\partial}{\partial \tau}\big(\xi_{u_ju_{j'}}^{(0)}(t,\tau)+\frac{1}{\gamma}\xi_{u_ju_{j'}}^{(1)}(t,\tau)+\ldots\big)\\
&-\sum_{\alpha,\alpha'=\pm 1}e^{i(\alpha+\alpha')\omega_0\tau}\bigg\{\sum_{\ell=\pm 1}\Big[A_{u_ju_{j'}}(\alpha,\alpha',\ell,t)\rho_{u_ju_\ell}
+B_{u_ju_{j'}}(\alpha,\alpha',\ell,t)\rho_{u_\ell u_{j'}}\Big]
+\sum_{\ell,\ell'=\pm 1}C_{u_ju_{j'}}(\alpha,\alpha',\ell,\ell',t)\rho_{u_{\ell'}u_\ell}\bigg\}\\
&-\frac{1}{\gamma}\sum_{\alpha,\alpha'=\pm 1}e^{i(\alpha+\alpha')\omega_0\tau}\bigg\{\sum_{\ell=\pm 1}\Big[A_{u_ju_{j'}}(\alpha,\alpha',\ell,t)
\big(\xi_{u_ju_\ell}^{(0)}(t,\tau)+\frac{1}{\gamma}\xi_{u_ju_\ell}^{(1)}(t,\tau)+\ldots\big)\\
&\quad\quad\quad\quad\quad\quad\quad\quad\quad\quad\quad\quad\;
+B_{u_ju_{j'}}(\alpha,\alpha',\ell,t)\big(\xi_{u_\ell u_{j'}}^{(0)}(t,\tau)+\frac{1}{\gamma}\xi_{u_\ell u_{j'}}^{(1)}(t,\tau)+\ldots\big)\Big]\\
&\quad\quad\quad\quad\quad\quad\quad\quad\quad\;\; +\sum_{\ell,\ell'=\pm 1}C_{u_ju_{j'}}(\alpha,\alpha',\ell,\ell',t)\big(\xi_{u_{\ell'}u_{\ell}}^{(0)}(t,\tau)+\frac{1}{\gamma}\xi_{u_{\ell'}u_\ell}^{(1)}(t,\tau)+\ldots\big)\bigg\}.
\end{align*}
Collecting the terms of order $1/\gamma^0$, $1/\gamma^1$, $1/\gamma^2, \ldots$ and setting each to zero produces a hierarchy of \LV{partial} DEs, which can be solved to obtain approximate solutions for $\rho_{u_ju_{j'}}(t)$ and $\xi_{u_ju_{j'}}(t,\tau)$. The lowest-order approximation is obtained by collecting the terms of order $1/\gamma^0$,
\begin{align*}
0=&\;\eta_{u_ju_{j'}}^{(0)}(t)+\frac{\partial\xi_{u_ju_{j'}}^{(0)}(t,\tau)}{\partial\tau}\;
-\sum_{\alpha,\alpha'=\pm 1}e^{i(\alpha+\alpha')\omega_0\tau}\bigg\{\sum_{\ell=\pm 1}\Big[A_{u_ju_{j'}}(\alpha,\alpha',\ell,t)\rho_{u_ju_\ell}^{(0)}+B_{u_ju_{j'}}(\alpha,\alpha',\ell,t)\rho_{u_\ell u_{j'}}^{(0)}\Big]\\&\quad\quad\quad\quad\quad\quad\quad\quad\quad\quad\quad\quad\quad\quad\quad
-\sum_{\ell,\ell'=\pm 1}C_{u_ju_{j'}}(\alpha,\alpha',\ell,\ell',t)\rho_{u_{\ell'}u_{\ell}}^{(0)}\bigg\}.
\end{align*}
Above, to make it explicit that we are considering the lowest order solution, we have taken $\rho_{u_ju_{j'}}(t)\rightarrow \rho_{u_ju_{j'}}^{(0)}(t)$.
Integrating this expression over $\tau$ yields 
\begin{align*}
0=&\;\xi_{u_ju_{j'}}^{(0)}(t,\tau)-\xi_{u_ju_{j'}}^{(0)}(t,0)+\eta_{u_ju_{j'}}^{(0)}(t)\tau\\&-\sum_{\alpha,\alpha'=\pm 1}
 \int_0^\tau d\tau'e^{i(\alpha+\alpha')\omega_0\tau'}  \bigg\{\sum_{\ell=\pm 1}\Big[A_{u_ju_{j'}}(\alpha,\alpha',\ell,t)\rho_{u_ju_\ell}^{(0)}+B_{u_ju_{j'}}(\alpha,\alpha',\ell,t)\rho_{u_\ell u_{j'}}^{(0)}\Big]\\&\quad\quad\quad\quad\quad\quad\quad\quad\quad\quad\quad\quad\quad\quad\quad
+\sum_{\ell,\ell'=\pm 1}C_{u_ju_{j'}}(\alpha,\alpha',\ell,\ell',t)\rho_{u_{\ell'}u_\ell}^{(0)}\bigg\}\\
=&\;\xi_{u_ju_{j'}}^{(0)}(t,\tau)-\xi_{u_ju_{j'}}^{(0)}(t,0)+\eta_{u_ju_{j'}}^{(0)}(t)\tau\;-\sum_{\alpha=\pm 1}
\tau\bigg\{\sum_{\ell=\pm 1}\Big[A_{u_ju_{j'}}(\alpha,-\alpha,\ell,t)\rho_{u_ju_\ell}^{(0)}+B_{u_ju_{j'}}(\alpha,-\alpha,\ell,t)\rho_{u_\ell u_{j'}}^{(0)}\Big]\\&\quad\quad\quad\quad\quad\quad\quad\quad\quad\quad\quad\quad\quad\quad\quad\quad\quad\quad\quad\quad\;
+\sum_{\ell,\ell'=\pm 1}C_{u_ju_{j'}}(\alpha,-\alpha,\ell,\ell',t)\rho_{u_{\ell'}u_\ell}^{(0)}\bigg\}\\
& +i\sum_{\substack{\alpha,\alpha'=\pm 1\\\alpha+\alpha'\neq0}}
\frac{e^{i(\alpha+\alpha')\omega_0\tau}-1}{(\alpha+\alpha')\omega_0}\bigg\{\sum_{\ell=\pm 1}\Big[A_{u_ju_{j'}}(\alpha,-\alpha,\ell,t)\rho_{u_ju_\ell}^{(0)}+B_{u_ju_{j'}}(\alpha,-\alpha,\ell,t)p_{u_\ell u_{j'}}^{(0)}\Big] \\&
\quad\quad\quad\quad\quad\quad\quad\quad\quad\quad\quad\quad\quad
+\sum_{\ell,\ell'=\pm 1}C_{u_ju_{j'}}(\alpha,-\alpha,\ell,\ell',t)\rho_{u_{\ell'}u_\ell}^{(0)}\bigg\}.
\end{align*}
In the last equality, we have grouped the terms with $\alpha+\alpha'=0$, which are proportional to $\tau$, and those with $\alpha+\alpha'\neq0$, which contain rapid oscillations in $\tau$. Observe that the ``secular terms" proportional to $\tau$ will grow without bound unless
\begin{eqnarray}
\dot\rho_{u_ju_{j'}}^{(0)}=\eta_{u_ju_{j'}}^{(0)}(t) 
& =& \sum_{\alpha=\pm 1}
\bigg\{\sum_{\ell=\pm 1}\Big[A_{u_ju_{j'}}(\alpha,-\alpha,\ell,t)\rho_{u_ju_\ell}^{(0)}+B_{u_ju_{j'}}(\alpha,-\alpha,\ell,t)\rho_{u_\ell u_{j'}}^{(0)}\Big] \notag \\
& + & \hspace*{10mm}\sum_{\ell,\ell'=\pm 1}C_{u_ju_{j'}}(\alpha,-\alpha,\ell,\ell',t)\rho_{u_{\ell'}u_\ell}^{(0)}\bigg\}.
\label{eq::solvable1}
\end{eqnarray}

Note that this expression is equivalent to \erf{eq::SecularApprox} in the main text. In the first line of this expression, we have used $\eta_{u_ju_{j'}}^{(0)}(t)=\partial \rho_{u_ju_{j'}}^{(0)}/\partial t=\dot\rho_{u_ju_{j'}}^{(0)}(t)$. These ``solvability conditions" for each $u_j$ and $u_{j'}$ constitute a closed set of DEs, which can be solved, in principle, to obtain the $\rho_{u_ju_{j'}}^{(0)}(t)$. Along with the initial condition $\xi_{u_ju_{j'}}^{(0)}(t,0)=0$, the solvability conditions also imply
\begin{align}
\label{eq::solvable2}
\xi_{u_ju_{j'}}^{(0)}(t,\tau)=&\,-i \sum_{\substack{\alpha,\alpha'=\pm 1\\\alpha+\alpha'\neq0}}
\frac{e^{i(\alpha+\alpha')\omega_0\tau}-1}{(\alpha+\alpha')\omega_0}\bigg\{\sum_{\ell=\pm 1}\Big[A_{u_ju_{j'}}(\alpha,-\alpha,\ell,t)\rho^{(0)}_{u_ju_\ell}+B_{u_ju_{j'}}(\alpha,-\alpha,\ell,t)\rho^{(0)}_{u_\ell u_{j'}}\Big] \notag \\
&\qquad
+\sum_{\ell,\ell'=\pm 1}C_{u_ju_{j'}}(\alpha,-\alpha,\ell,\ell',t)\rho^{(0)}_{u_{\ell'}u_\ell}\bigg\}.
\end{align}
Eqs.\,(\ref{eq::solvable1}) and (\ref{eq::solvable2}) determine the first-order solution to the Ansatz in \erf{eq::ansatz}, $\rho_{u_ju_{j'}}(t,\tau)\approx \rho^{(0)}_{u_ju_{j'}}(t)+ \tfrac{1}{\gamma} \xi_{u_ju_{j'}}^{(0)}(t,\tau)$. Since $A_{u_ju_{j'}}(\alpha,\alpha',\ell,t),\,B_{u_ju_{j'}}(\alpha,\alpha',\ell,t),\,C_{u_ju_{j'}}(\alpha,\alpha',\ell,t)\sim t$, we must have $\xi_{u_ju_{j'}}^{(0)}(t,\tau)\sim t$ and $\rho^{(0)}_{u_ju_{j'}}(t)\sim t^2$. This implies that $\rho_{u_ju_{j'}}(t,\tau)\approx\rho^{(0)}_{u_ju_{j'}}(t)\gg\xi_{u_ju_{j'}}^{(0)}(t,\tau)/\gamma$ when $\omega_0\gamma t=\omega t\gg 1$. This recovers the lowest-order secular approximation we have employed.

\subsection{Specialization to single-axis dephasing noise}
\label{mesolve1axis}

Dynamics under single-axis noise can be recovered, of course, as a special case of the multi-axis dynamics. In the absence of transverse noise ($S_{-1,1}(\omega)=S_{1,-1}(\omega)=0$), only the dephasing spectrum is relevant, and $S_{0,0}(\omega)=S(\omega)$, as in the main text. Under a constant $x$-axis drive, the control matrix for such a system reads
\begin{align}
    y_{\alpha\beta}(t)=\begin{bmatrix}
        \cos^2\left(\frac{\Omega}{2}t\right) & i\frac{\sin\left(\Omega t\right)}{\sqrt{2}} & \sin^2\left(\frac{\Omega}{2}t\right)\\
        i\frac{\sin\left(\Omega t\right)}{\sqrt{2}} & \cos\left(\Omega t\right) & -i\frac{\sin\left(\Omega t\right)}{\sqrt{2}}\\
        \sin^2\left(\frac{\Omega}{2}t\right) & -i\frac{\sin\left(\Omega t\right)}{\sqrt{2}} & \cos^2\left(\frac{\Omega}{2}t\right)
\end{bmatrix} .
\end{align} 
Integrating over time in Eq.\,\eqref{eq::SecularApprox}, we obtain the following DE:
\begin{eqnarray}
    \dot{\rho}_{x_+x_+}(t)=\frac{i}{2\pi}\int_\infty^\infty d\omega\bigg\{\rho_{x_+x_+}(t)\Big[ 
    \left(\frac{e^{i(\omega-\Omega)t}-1}{\omega-\Omega}\right)S(-\omega)
    +\left(\frac{e^{i(\omega+\Omega)t}-1}{\omega+\Omega}\right)S(\omega) \Big]
    - \notag\\
    \rho_{x_-x_-}(t)\Big[\left(\frac{e^{i(\omega+\Omega)t}-1}{\omega+\Omega}\right)S(-\omega)
    +\left(\frac{e^{i(\omega-\Omega)t}-1}{\omega-\Omega}\right)S(\omega) \Big]\bigg\}, 
    \label{de1}
\end{eqnarray}
\vspace*{-5mm}
\begin{eqnarray}
    \dot\rho_{x_+x_-}(t)=-\frac{i}{2\pi}\int_{-\infty}^{\infty}d\omega\bigg\{\rho _{x_-x_+}(t)\left(\frac{e^{2 i\Omega t}-e^{i (\omega +\Omega )t}}{\Omega -\omega }\right)S^+(\omega)+\rho _{x_+x_-}(t)\left(\frac{1-e^{i (\omega +\Omega )t}}{\omega +\Omega }\right)S^+(\omega)\bigg\},
    \label{de2}
\end{eqnarray}
\end{widetext}
To derive the ME in Eqs.\,\eqref{slqnssym1}-\eqref{slqnssym2} in the main text, the secular term proportional to $e^{2i\Omega t}$ is considered to be rapidly oscillating under the integral, and thus ignored, whereas terms proportional to $e^{i(\omega\pm \Omega)t}$ are retained. Furthermore, if the evolution time is long but finite, $|\Omega| t\gg1$, the sinc functions in the integrand can be approximated to Dirac delta functions, 
$$\frac{\sin[(\omega\pm\Omega)t]}{\omega\pm\Omega} \approx \pi\delta(\omega\pm\Omega).$$
The sinc functions can be thought of as filter functions in the formalism of \cite{PazSilva2016}. They are centered at the driving amplitude $\Omega$ and sample the spectrum at $S(\Omega)$. In principle, this allows us to measure the qubit response to the spectrum at any frequency, as long as the condition $|\Omega| t\gg 1$ holds. Under this approximation, the desired toggling-frame ME is obtained.

To justify the secular approximation invoked above, let us retain all the oscillating terms in Eqs.\,\eqref{de1}-\eqref{de2}, and examine the dynamics in the rotating frame, where these equations become 
\begin{widetext}
\begin{eqnarray*}
    \dot{\rho}_{\text{rot }x_+x_+}(t)=\frac{i}{2\pi}\int_\infty^\infty d\omega\bigg\{\rho_{\text{rot }x_+x_+}(t)\Big[ 
    \left(\frac{e^{i(\omega-\Omega)t}-1}{\omega-\Omega}\right)S(-\omega)
    +\left(\frac{e^{i(\omega+\Omega)t}-1}{\omega+\Omega}\right)S(\omega) \Big]
    - \notag\\
    \rho_{\text{rot }x_-x_-}(t)\Big[\left(\frac{e^{i(\omega+\Omega)t}-1}{\omega+\Omega}\right)S(-\omega)
    +\left(\frac{e^{i(\omega-\Omega)t}-1}{\omega-\Omega}\right)S(\omega) \Big]\bigg\}, 
\end{eqnarray*}
\vspace*{-5mm}
\begin{eqnarray*}
    \dot\rho_{\text{rot }x_+x_-}(t) =-\frac{i}{2\pi}\int_{-\infty}^{\infty}d\omega\bigg\{\rho _{\text{rot }x_-x_+}(t)\left(\frac{1-e^{i (\omega -\Omega )t}}{\Omega -\omega }\right)S^+(\omega)+\rho _{\text{rot }x_+x_-}(t)\left(\frac{1-e^{i (\omega +\Omega )t}}{\omega +\Omega }\right)S^+(\omega)\bigg\} 
    -i\Omega \rho_{\text{rot }x_+x_-}(t).
\end{eqnarray*}
\end{widetext}
Invoking the delta-approximation in the filter function now yield the following rotating-frame ME:
\begin{align*}
    \dot{\rho}_{\text{rot }x_+x_+}(t)&=-\rho_{\text{rot }x_+x_+}(t)S(-\Omega)+\rho_{\text{rot }x_-x_-}(t)S(\Omega),\\
    \dot{\rho}_{\text{rot }x_+x_-}(t)&=-\frac{S^+(\Omega)}{2}\left[\rho_{\text{rot }x_+x_-}(t)-\rho_{\text{rot }x_-x_+}\right]\notag\\
    &\qquad\qquad\qquad\qquad-i\Omega\rho_{\text{rot }x_+x_-}(t).
\end{align*}
The secular term in $\Omega$ does {\em not} appear in the rotating-frame equations, even though it has not been discarded at the time. Solving for the coherence dynamics, we find
\begin{eqnarray*}
&\rho _{\text{rot }x_+x_-}&\hspace*{-2mm}(t)=e^{-\frac{1}{2}S^+(\Omega)t} \bigg[ \rho _{\text{rot }x_+x_-}(0) \cos(\Omega \Delta t) \\
&+\hspace*{-3mm}& \!\!\!\!\!\! \frac{i \sin(\Omega\Delta t)}{\Delta}
\Big( \rho _{\text{rot }x_-x_+}(0)\frac{S^+(\Omega)}{2i\Omega}-\rho_{\text{rot }x_+x_-}(0) \Big) \bigg],
\end{eqnarray*}
where $\Delta \equiv \sqrt{1- \tfrac{1}{4\Omega^2} S^+(\Omega)^2}$. 
This shows that, by retaining the secular terms in toggling-frame, the rotating-frame dynamics couple the coherence terms with a term proportional to the ratio $\frac{S^+(\Omega)}{\Omega}$. Thus, for the coherence dynamics to be well-approximated by the secular approximation, this coupling term must be very small. In QNS settings as we consider, the driving amplitude is typically significantly larger than the native noise, ensuring that $$  \left|\frac{S^+(\Omega)}{\Omega}\right|\ll 1,$$ also meaning that $\Delta \approx 1$. Therefore, dropping terms proportional to this ratio is akin to making the secular approximation in the toggling-frame. 

\section{SPAM characterization with general initialization and readout}
\label{app:sp}

The SP described in Eq.\,\eqref{statepreperr} can represent \emph{any} state once a specific axis $u$ has been chosen. Selecting $u=x$, one can think of the SP parameters $\alpha_{SP}$, $\text{Re}(c_x)$ and $\text{Im}(c_x)$ as the components of a Bloch vector in the basis $\{\ket{x_+}, \ket{x_-}\}$. Therefore, without placing any constraints on these parameters, we can consider the most general SP in the SL QNS setting by inserting the general initial state $\varrho_{\ket{x_\pm}}(0)$ into Eq.\,\eqref{eq::ExpectX},
\begin{align}
    \widehat{\braket{\sigma_x(T)}}_{\varrho_{\ket{x_\pm}}(0)}^{\Omega_x} &= \alpha_M \frac{S^-(\Omega)}{S^+(\Omega)}\left(1-e^{-S^+(\Omega)T}\right) \notag\\
    &+ \alpha_M\Big\{\left[\varrho_{\ket{x_\pm}}(0)\right]_{x_+x_+}\notag\\
    &- \left[\varrho_{\ket{x_\pm}}(0)\right]_{x_-x_-}\Big\} e^{-S^+(\Omega)T} + \delta,
\end{align}
which leads to $\left[\varrho_{\ket{x_\pm}}(0)\right]_{x_+x_+} - \left[\varrho_{\ket{x_\pm}}(0)\right]_{x_-x_-}=\pm\alpha_{SP}$, and the resulting equation for the observable expectation value
\begin{eqnarray}
    \widehat{\braket{\sigma_x(T)}}_{\varrho_{\ket{x_\pm}}(0)}^{\Omega_x} &=&\alpha_M \frac{S^-(\Omega)}{S^+(\Omega)}\left(1-e^{-S^+(\Omega)T}\right) \notag\\
    &\pm &\alpha e^{-S^+(\Omega)T} + \delta,
\end{eqnarray}
is completely determined through the parameters $\alpha$, $S^+(\Omega)$, $\delta$, $\alpha_M$ and $S^-(\Omega)$. In the main text, we showed that $S^+(\Omega)$, $\alpha$ and $\delta$ can be estimated independently but not $\alpha_M S^-(\Omega)$. To be sure that there is no measurement on the final state that can deconvolve these two parameters, consider the other observable expectation values keeping the faulty initial conditions the same (at times $T^{(n)}_j=2\pi n_j/|\Omega|$, for some $n_j\in \mathbb{N}$, if needed to ensure frame alignment),
\begin{align*}
    \widehat{\braket{\sigma_y(T^{(n)}_j)}}_{\varrho_{\ket{x_\pm}}(0)}^{\Omega_x}&=\mp\alpha_M\text{Im}(c_x) e^{-\frac{S^+(\Omega)}{2}T^{(n)}_j}+\delta,\\
    \widehat{\braket{\sigma_z(T^{(n)}_j)}}_{\varrho_{\ket{x_\pm}}(0)}^{\Omega_x}&=\pm\alpha_M\text{Re}(c_x) e^{-\frac{S^+(\Omega)}{2}T^{(n)}_j}+\delta.
\end{align*}
Together, these expectation values completely characterize the final state that is measured. The expectation values perpendicular to the drive introduce new products of $\alpha_M$ with $\text{Re}(c_x)$ and $\text{Im}(c_x)$.

Similarly, the steady-state dynamics also do not provide any new relationships between the parameters because,
\begin{align*}
    \widehat{\braket{\sigma_x}}_\text{ss}^{\Omega_x}&=\alpha_M\frac{S^-(\Omega)}{S^+(\Omega)}+\delta, \\
    \widehat{\braket{\sigma_z}}^{\Omega_x}_\text{ss}&=\widehat{\braket{\sigma_y}}^{\Omega_x}_\text{ss}=\delta.
\end{align*}
Therefore, from just the final state, it is {\it not} possible to deconvolve $\alpha_M$ and $S^-(\Omega)$. If, however, $\alpha_{SP}$ is characterized independently from a separate protocol then $\alpha_M$ can be characterized and hence the SPAM free quantum spectrum $S^-(\Omega)$.

\section{Numerical simulation of temporally correlated non-classical dephasing}
\label{quantnoisesim}

We use discrete spectral analysis (DSA), to numerically generate classical Gaussian noise with a prescribed power spectral density \cite{kozachenko_simulation_1994}.

The goal is to generate a noise vector $\beta(t)$ that describes a noise trajectory in time, generated from a zero-mean, stationary Gaussian process with a spectrum $\Tilde{S}(\omega)$ that we call the ``generating spectrum''. We choose $\Tilde{S}(\omega)$ such that it is an even function and decays $\Tilde{S}(\omega_\text{max})\approx 0$ to zero beyond some cutoff frequency $\omega_\text{max}$. We can then discretize the frequency interval $[0,\omega_\text{max}]$ into $n_\omega$ intervals of $d\omega \equiv \omega_\text{max}/n_\omega$. Then we let $\{A_1,\ldots, A_{n_\omega}\}$ and $\{B_1,\ldots, B_{n_\omega}\}$ be independent, zero-mean Gaussian white-noise processes such that,
\begin{align*}
    \braket{A_iA_j} = \delta_{ij}, \quad
    \braket{B_iB_j} = \delta_{ij},\quad 
    \braket{A_iB_j} = 0.
\end{align*}
By passing these uncorrelated random variables through a non-linear filter, we claim that
\begin{align}
    \beta(t_i) \equiv  \sum^{n_\omega-1}_{j=0}G_j[A_j\cos(\omega_j t_i)+B_j\sin(\omega_j t_i)], 
    \label{eqbeta}
\end{align}
with $\omega_j=jd\omega$, is a trajectory generated from a stationary Gaussian random process, where $G_j = \sqrt{{d \omega \Tilde{S}(\omega_j)}/{\pi}}$.

We can check our claim by calculating the correlation function while assuming time to be continuous,
\begin{align}
\label{corrftn}
    \braket{\beta(t)\beta(s)} &= \sum^{n_\omega-1}_{j=0}G_j^2\cos(\omega_j(t-s)).
\end{align}
As shown, the noise process is stationary and one can check that all higher-order correlations are zero and, since $\beta(t)$ in Eq.\,\eqref{eqbeta} is zero-mean, the process is Gaussian. However, this process only lets us generate a noise trajectory, and on its own it can only represent a classical temporally correlated noise process given by the dephasing system-bath Hamiltonian $
    H_{SB}(t)=\sigma_z\otimes \beta(t)\mathbb{1}_B.$ Since SL QNS can estimate both classical and quantum spectra, we want to use these noisy non-Markovian trajectories to simulate non-commuting bath operators to validate our single-axis SPAM-robust protocol.

To model a qubit in an environment where the bath operators do not commute in time, the simplest toy model is a two-qubit system. The most general coupling Hamiltonian for such a system can be written as
\begin{align}
\label{simnoise}
    H_{SB}(t) = \sigma_z\otimes\left[\frac{\beta_x(t)}{2}\tau_x+\frac{\beta_y(t)}{2}\tau_y+\frac{\beta_z(t)}{2}\tau_z\right], 
\end{align}
where Pauli operators $\sigma_i$ ($\tau_i$) act on the system (bath) qubit. For simplicity, the bath Hamiltonian can be chosen as $H_B=0$ thereby ensuring that any choice of the initial bath state satisfies the stationary condition $[H_B,\rho_B(0)]=0$. The bath operator is thus
\begin{align*}
    B(t) &= \frac{\beta_x(t)}{2}\tau_x+\frac{\beta_y(t)}{2}\tau_y+\frac{\beta_z(t)}{2}\tau_z. 
\end{align*}
Direct calculation shows that the anti-commutator reads
\begin{multline*}
    \{B(t_1),B(t_2)\} = \frac{1}{2}[\beta_x(t_1)\beta_x(t_2)+\beta_y(t_1)\beta_y(t_2)\\
    +\beta_z(t_1)\beta_z(t_2)],
\end{multline*}
whereas for the commutator we have 
\begin{multline*}
    [B(t_1),B(t_2)] = \frac{i}{2}[(\beta_x(t_1)\beta_y(t_2)-\beta_y(t_1)\beta_x(t_2))\tau_z\\
    +(\beta_y(t_1)\beta_z(t_2)-\beta_z(t_1)\beta_y(t_2))\tau_x\\
    +(\beta_z(t_1)\beta_x(t_2)-\beta_x(t_1)\beta_z(t_2))\tau_y].
\end{multline*}
If $\beta_x(t)=\beta_y(t)=\beta_z(t)$ are identical, the commutator vanishes and the noise model is symmetric. However, let us, for instance, choose $\rho_B(0)=\ket{0}\bra{0}_B$ and set
\begin{align}
\label{timelag}
    \beta_x(t) =\beta_z(t) = \beta(t), \quad 
    \beta_y(t) = \beta(t+\gamma), 
\end{align}
with $\gamma$ as a time-lag. Then by setting $t_1-t_2\equiv \tau$, the above quantum and classical expectation values become
\begin{align*}
    \braket{\{B(\tau),B(0)\}}   &= \frac{3}{2}\sum^{n_\omega-1}_{j=0}\!G_j^2\cos(\omega_j\tau),
\end{align*}
\vspace*{-5mm}
$$    \braket{[B(t_1),B(t_2)]}  \! = \!\frac{i}{2}\sum^{n_\omega-1}_{j=0}\!\!G_j^2[\cos(\omega_j(\tau-\gamma))\\-\cos(\omega_j(\tau+\gamma))].$$

The classical and quantum spectra generated from this model can be obtained through a Fourier transform of the time-correlation functions. The classical spectrum is simply a re-scaled form of the generating spectrum, $S^+(\omega)=\tfrac{3}{2}\Tilde{S}(\omega),$ and since the latter is symmetric, $S^+(\omega)$ is a symmetric function as well. Similarly, the quantum spectrum reads $S^-(\omega)=\Tilde{S}(\omega)\sin(\gamma\omega),$
which, as expected, it is anti-symmetric about the $\omega=0$. If, in Eq.\,\eqref{timelag}, we let $\beta_z(t)=0$ as in the main text, then $S^+(\omega)=\tilde S(\omega)$ and $S^-(\omega)=\tilde S(\omega)\sin(\gamma\omega)$.

\section{Details of implementation and data analysis}
\label{app:expt}

\subsection{Implementing a constant drive}
\label{app::constdrive}

The first step in implementing the protocol is the calibration of pulse amplitudes. Since the Qiskit-pulse API works in the qubit co-rotating frame, the pulses one submits are constant drive pulses.

Consider the constant $x$-drive pulses. While the pulse profiles needed are built in, we chose the driving amplitude $\Omega$ to sample the spectra at the desired frequencies. The API only allows for submitting amplitudes on an undefined unit scale $-1\leq \Omega_\text{IBM} \leq 1$, where $\Omega_\text{IBM}$ is the choice of $\Omega$ in the undefined units. To ensure proper calibration, we chose a set of amplitudes $\{\Omega_\text{IBM}^i\}$ and implement Rabi experiments for each $\Omega_\text{IBM}^i$. The transition probabilities in the computational basis, ${\mathbb P}_{\ket{z_+}\rightarrow\ket{z_-}}$ were measured and fit to a sinusoidal function. The estimated parameter to fit was $\Omega$ in the proper units (Hz). This generated a map $\{\Omega_\text{IBM}^i\}\rightarrow\{\Omega_i\}$ that we utilized in our SL QNS protocol.

Another limitation of the API is the number of pulse samples that can be sent in one job. One job may only have at most $258,144$ samples, where each sample is $0.22ns$. This amounts to a pulse that is only $56.8\mu s$ in length. Since our  experiments require multiple runs with pulses of length $\sim 20\mu s$, a new approach was required. This was done by building custom gates that generated a constant $x$-drive for $19\mu s$ and concatenating them.  Doing so bypassed the API limitation successfully, at the cost of introducing some noise due to the ramp-up and ramp-down of the concatenated pulses.

Now consider the constant on-axis $z$-drive. Qiskit implements direct rotations along the $z$-axis $R_z(\theta)$ virtually \cite{PhysRevA.96.022330} through a phase-shift, limiting out ability to directly generate $z$-drive control Hamiltonians. To generate a continuous drive using these gates, we interleave an instantaneous $R_z(\delta\theta)$ with small time steps of free evolution. The descretized model can be related to the continuous driving by defining $U_\mathrm{f}(t)$ to be the free-evolution under the noisy-bath Hamiltonian $H_{SB}(t)$,
\begin{align*}
    U_\mathrm{f}(t_1,t_2)&={\cal T}_+\exp \Big\{-i\int_{t_1}^{t_2}H_{SB}(s)ds\Big\}.
\end{align*}
Let $U_\mathrm{tot}(t)$ be the evolution under both a continuous $z$-drive Hamiltonian $H_\text{ctrl}(t)$ and $H_{SB}(t)$,
\begin{align*}
    U_\mathrm{tot}(T)&={\cal T}_+\exp \Big\{-i\int_{0}^{T}[H_\text{ctrl}(s)+H_{SB}(s)]ds\Big\}.
\end{align*}
The discretized model corresponding to this unitary evolution by dividing the total evolution time T into steps of $\delta t=T/N$. The corresponding unitary with $\delta \theta = \Omega\delta t/2$ is
\begin{align*}
    U_\mathrm{eff}(T) &= \prod_{i=0}^{N-1} R_z(\tfrac{\Omega}{2}\delta t)U_f(t_i, t_i+\delta t).
\end{align*}
In comparison consider the exact unitary corresponding to the small time-step $\delta t$,
\begin{align*}
    U_\mathrm{tot}(t+\delta t, t) &= e^ {-i [\tfrac{\Omega}{2}\sigma_z+\sum_\alpha \sigma_\alpha B_{-\alpha}(t)]\delta t} .
\end{align*}
Assuming that the noise operators vary slow enough so that $B_\alpha(t)$ are constant between $t$ and $t+\delta t$, a `Trotterization' shows that the effective unitary is equal to the actual unitary up to the first order, by applying the BCH formula to $U_\mathrm{eff}(\delta t)$,
$$ U_\mathrm{eff}(t+\delta t,t) = 
    e^{-i [\tfrac{\Omega}{2}\sigma_z+\sum_{\alpha}\sigma_\alpha B_{-\alpha}(t)]\delta t -iO(\delta t^2)}.$$
Thus, $U_\mathrm{eff}(\delta t)\approx U_\mathrm{tot}(\delta t)$ for a small enough time-step. We chose $N=10^3$, $\delta t^2/T^2= 10^{-6}$, after trial and error and verification through numerical simulations, with the aim of minimizing the number of gates needed while keeping the interval as short as possible. If a user has access to more experimental resources, $\delta t$ can be made smaller.

\subsection{Frame-alignment and data-points for linear-regression}
\label{app::framealign}

The theoretical framework of SL QNS is formulated in the toggling-frame with respect to the control Hamiltonian $H_\text{ctrl}(t)$. However, measurements on qubits are obtained in the qubit co-rotating frame. While most observable expectation values relevant to SL QNS are identical in the rotating-frame and the toggling-frame, there are some observables that are oscillatory in the rotating-frame, yet are monotonically decreasing function in the toggling-frame as in Eq.\,\eqref{eq::togsxsol}. Therefore, for some observables it is useful to select qubit evolution times such that the rotating and the toggling-frames align. As we discussed in the main text, this can be achieved by ensuring the condition $T^{(n)}_j=2\pi n_j/|\Omega_i|$. Experimental systems limit us in such a way that we can select times that satisfy this condition and not the amplitude $\Omega$. The calibrated set of amplitudes $\{\Omega_i\}$ is used to find a corresponding set of times $\{T^{(n)}_{j}=2\pi n_{j}/|\Omega_i|\}$, $n_{j}\in\mathbb{N}$. These evolution times must satisfy the long-time condition $T^{(n)}_{j}|\Omega_i|\gg 1$ as well as the frame-alignment condition, and $n_{i0}<n_{i1}<\ldots <n_{i(M-1)}$. $M$ represents the number of data-points acquired in the time series analysis for each amplitude $\Omega_i$. With this choice of time the complete set of expectation values can be obtained. It is important to note that this condition is only needed to estimate observables that depend on the coherence of the qubit $\rho_{u_\pm u\mp}(t)$ under a drive along the $u$-axis. When the frames are aligned, $\rho_{u_\pm u\mp}(T^{(n)}_{j})=\rho_{\text{rot, }u_\pm u\mp}(T^{(n)}_{j})$.

\subsection{Error analysis}

In analyzing data from SL experiments, we consider two primary sources of uncertainty: i) finite measurement statistics; and ii) weighted linear regression. Finite measurement statistics arise from the necessarily limited number of shots that can be taken experimentally. Under $N_\mathrm{shots}$ trials, the estimated expectation values have a variance
\begin{align*}
    \mathrm{Var}(\widehat{\braket{\sigma_x(T_{j})}}^{\Omega_i}_{x_\pm}) &= \frac{{\mathbb P}(x_+, \rho(T_{j})) {\mathbb P}(x_-, \rho(T_{j}))}{N_\mathrm{shots}},\\
    \mathrm{Var}(\widehat{\braket{\sigma_z(T_{j})}}^{\Omega_i}_{z_\pm}) &= \frac{{\mathbb P}(z_+, \rho(T_{j})) {\mathbb P}(z_-, \rho(T_{j}))}{N_\mathrm{shots}}.
\end{align*}
The corresponding standard deviation is used as the standard error in the estimated expectation values. Under the assumption that $N_\mathrm{shots}\gg 1$, the binomial distribution that is associated with the measurement outcomes can be approximated to a Gaussian distribution. With $M$ data-points in time for each $\Omega_i$, the left-hand sides of the equations in Sec.\,\ref{sec::multiqnsprotocol} are evaluated and the standard errors propagated. Since we are using the experimental data as inputs to non-linear functions, we use a weighted linear-regression on $M$ data-points. The covariance matrix generated in this process is used to get the standard error in the estimated slope and intercept. For our experiments $N_\mathrm{shots}=2,000$ and $M = 10$.

%

\end{document}